\pdfoutput=1
\documentclass[twocolumn]{article}

\usepackage[modulo,switch]{lineno}
\modulolinenumbers[1]

\usepackage[utf8]{inputenc}
\usepackage[T1]{fontenc}
\usepackage{hyperref}
\usepackage{amsmath}
\usepackage{amssymb}
\usepackage{siunitx}
\usepackage{graphicx}
\usepackage{xcolor}
\usepackage{stmaryrd}
\usepackage{empheq}

\makeatletter\@ifundefined{date}{}{\date{}}
\makeatother

\evensidemargin\oddsidemargin \hoffset-22.4mm \voffset-28.4mm
\newcommand{\nin}{\noindent}

\newcommand{\pr}[1]{\left( #1 \right)}
\newcommand{\hk}[1]{\left[ #1 \right]}
\newcommand{\dhk}[1]{\llbracket #1 \rrbracket}
\newcommand{\br}[1]{\left\{ #1 \right\}}
\newcommand{\md}[1]{\left| #1 \right|}

\newcommand{\gu}[1]{`#1'}

\newcommand{\R}{\mathbb{R}}
\newcommand{\C}{\mathbb{C}}

\newcommand{\w}{\omega}

\newcommand{\Ze}{Z_{\text{exp}}}
\newcommand{\Zr}{Z_{\text{fit}}}
\newcommand{\wl}{\omega_\ell}
\newcommand{\thl}{\theta_\ell}
\newcommand{\fl}{f_\ell}
\newcommand{\Ql}{Q_\ell}
\newcommand{\gaml}{\frac{\wl}{\Ql}}
\newcommand{\agaml}{\frac{\thl}{\Ql}}

\newcommand{\fs}{f_{\text{thresh}}}
\newcommand{\pso}[1]{p_{\text{thresh},n}^{\text{opt,#1}}}
\newcommand{\fso}[1]{f_{\text{thresh},n}^{\text{opt,#1}}}
\newcommand{\flo}[1]{f_{\ell,n}^{\text{opt,#1}}}
\newcommand{\pson}[2]{p_{\text{thresh},#2}^{\text{opt,#1}}}
\newcommand{\fson}[2]{f_{\text{thresh},#2}^{\text{opt,#1}}}
\newcommand{\flon}[2]{f_{\ell,#2}^{\text{opt,#1}}}

\newcommand{\wlql}{\frac{\wl}{\Ql}}

\newcommand{\ad}[1]{\widetilde{#1}}

\newcommand{\hp}{\dot{h}}

\newcommand{\kap}[1]{\frac{C_{#1}}{\w_0}}
\newcommand{\sig}[1]{\frac{s_{#1}}{\w_0}}
\newcommand{\X}{\mathbf{X}}

\renewcommand{\Re}{\mathfrak{Re}}
\renewcommand{\Im}{\mathfrak{Im}}

\newcommand{\J}{\mathrm{j}}
\newcommand{\e}{\mathrm{e}}
\DeclareMathOperator{\sgn}{\text{sgn}}

\newcommand{\iTF}[1]{\mathfrak{F}^{-1}\hk{#1}}
\newcommand{\diff}{\mathop{}\!\mathrm{d}}

\begin{document}

\title{Minimal blowing pressure allowing periodic oscillations in a model of bass brass instruments}

\author{Rémi Mattéoli$^{*}$, Joël Gilbert$^{*}$, Christophe Vergez$^{\dagger}$,\\ Jean-Pierre Dalmont$^{*}$, Sylvain Maugeais$^{\ddagger}$, Soizic Terrien$^{*}$, Frédéric Ablitzer$^{*}$ \\
$^{*}$ Laboratoire d'Acoustique de l'Université du Mans (LAUM), UMR CNRS 6613,\\ Institut d'Acoustique - Graduate School (IA-GS), CNRS, Le Mans Université, France\\
$^{\dagger}$ Aix Marseille Univ, CNRS, Centrale Marseille, LMA, UMR 7031, France\\
$^{\ddagger}$ Laboratoire Manceau de Mathématiques – Le Mans Université, 72085 Le Mans, France}

\maketitle\thispagestyle{empty}

\begin{abstract}

In this study, an acoustic resonator -- a bass brass instrument -- with multiple resonances coupled to an exciter -- the player's lips -- with one resonance is modelled by a multidimensional dynamical system, and studied using a continuation and bifurcation software. Bifurcation diagrams are explored with respect to the blowing pressure, in particular with focus on the minimal blowing pressure allowing stable periodic oscillations and the associated frequency.

The behaviour of the instrument is first studied close to a (non oscillating) equilibrium using linear stability analysis. This allows to determine the conditions at which an equilibrium destabilises and as such where oscillating regimes can emerge (corresponding to a sound production). This approach is useful to characterise the ease of playing of a brass instrument, which is assumed here to be related -- as a first approximation -- to the linear threshold pressure. In particular, the lower the threshold pressure, the lower the physical effort the player has to make to play a note \cite{campbellScienceBrassInstruments2021}.

Cases are highlighted where periodic solutions in the bifurcation diagrams are reached for blowing pressures below the value given by the linear stability analysis. Thus, bifurcation diagrams allow a more in-depth analysis. Particular attention is devoted to the first playing regime of bass brass instruments (the pedal note and the ghost note of a tuba in particular), whose behaviour qualitatively differs from a trombone to a euphonium for instance.

\end{abstract}

\section{Introduction}

One main goal of the acoustics of wind instruments is to describe and quantify the intonation and ease of playing of an instrument. From the physics viewpoint, it is interesting to model the coupled system formed by the musician and the instrument. Of particular interest is the influence of the musician's control parameters on the oscillation frequency (linked to the intonation), and the minimum mouth pressure required to achieve auto-oscillations (related to the ease of playing). Indeed, it is assumed here that the musician's feeling of ease of playing partly relies on the threshold blowing pressure: the higher the latter, the higher the physical effort the player has to make to play a note \cite{campbellScienceBrassInstruments2021}. In practice, the musician can play several notes without depressing any valves in the case of a tuba, or moving the slide in the case of a trombone. These playing regimes are called the natural notes (B$\flat$1, B$\flat$2, F3, B$\flat$3, D4, F4,... in the case of a trombone or a euphonium for instance), and their frequencies are close to the resonance frequencies of the instrument as a whole, except for the lowest note playable (B$\flat$1).

Several tools are available to infer the oscillation frequencies and the threshold blowing pressures from a model of the system. Linear stability analysis allows to predict the behaviour of a dynamical system in the vicinity of its equilibrium solutions. The method consists in linearising the system and analysing its eigenvalues to determine whether or not the equilibrium solutions are stable. This method has already been applied to physical models of musical instruments -- brass instruments -- in \cite{velutHowWellCan2017}. Linear stability analysis is, in essence, only valid close to the equilibrium solutions. As such, this method alone might not explain the existence of periodic regimes, as highlighted for instance in \cite{velutHowWellCan2017} for the first regime of a saxhorn.

Alternatively, bifurcation analysis gives access to an extensive knowledge of permanent regimes (namely equilibrium and periodic regimes) of the dynamical system, as well as their stability when relevant. Bifurcation diagrams ideally represent all the families of equilibrium and periodic solutions with respect to one parameter of interest. In this paper, we investigate bifurcation diagrams of a physical model of a brass instrument to understand some aspects of its behaviour.

We consider here a simple model: the player’s lips are modeled through the \gu{outward-striking valve} \cite{campbellScienceBrassInstruments2021,cullenBrassInstrumentsLinear2000,elliottRegenerationBrassWind1982}, a one degree-of-freedom system. Also, the nonlinear sound propagation in the instrument's bore which accounts for the \gu{brassy sounds} \cite{campbellScienceBrassInstruments2021} at high sound levels is neglected \cite{msallamPhysicalModelTrombone2000,berjaminTimedomainNumericalModeling2017}. The airflow blown by the musician into the instrument is described by a nonlinear algebraic equation. Some of the functions in this equation being non-smooth, a regularised form of this equation is used for numerical reasons.

In section \ref{TB}, we describe the considered model and numerical methods, namely linear stability analysis and numerical continuation. In sections \ref{LPLSA} and \ref{BDMBP}, we use linear stability analysis and bifurcation diagrams, respectively, to investigate the minimal blowing pressure and playing frequencies of the trombone. Eventually, we focus in section \ref{FR}  on the first regime of two bass brass instruments (the trombone and the euphonium), mainly thanks to bifurcation diagrams. This highlights a major difference in the behaviour of these two instruments.

\section{Theoretical background}
\label{TB}

\subsection{Generic brass model}
\label{GBM}

This subsection details the brass-instrument model considered throughout the article. 
Brass instruments as a whole can be described through both linear and nonlinear mechanisms. More precisely, a localised nonlinear element (the lips' valve effect, i.e. the airflow modulation caused by the lips' vibration) excites a passive linear acoustic multimode element (the musical instrument, usually characterised by its input impedance in the frequency domain). The latter acoustic resonator exerts, in turn, a retroaction on the former mechanical resonator. Such musical instruments are self-sustained oscillators: they generate an oscillating acoustic pressure (the note played) from a static overpressure in the player's mouth (the blowing pressure).\\

\begin{figure}[h!]
    \centering
    \includegraphics[width=0.6\linewidth]{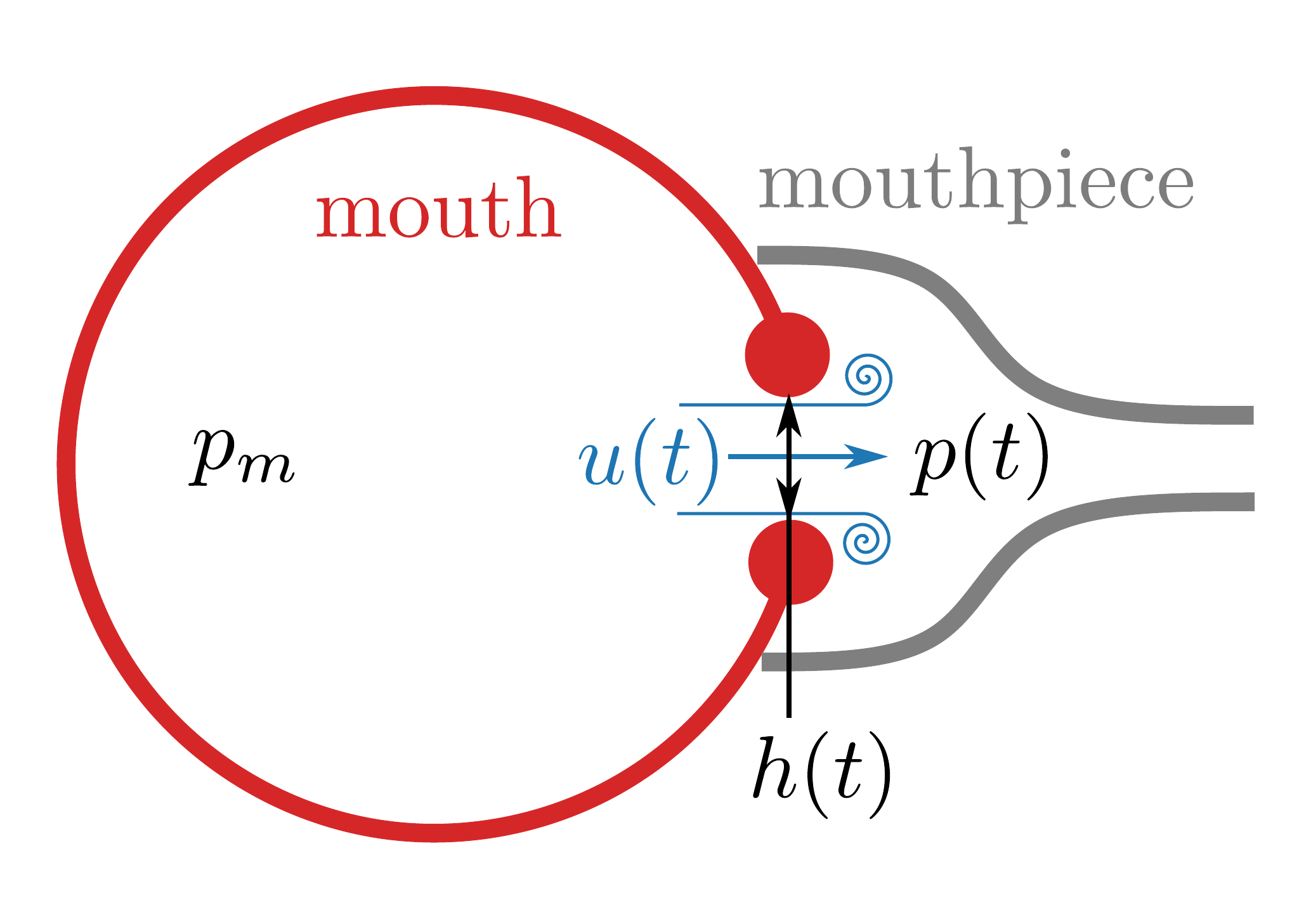}
    \caption{\baselineskip 3mm Schematic representation of the player's mouth coupled to the mouthpiece of a brass instrument.}
    \vspace{0mm}
    \label{schema_mouthpiece}
\end{figure}

The brass instrument coupled to the player is then described by a system of three equations. Because it relies on major simplifications \cite{campbellScienceBrassInstruments2021,elliottRegenerationBrassWind1982}, this model is often referred to as \gu{elementary}. More precisely, the three equations link the lip-opening height $h(t)$, the pressure in the mouthpiece $p(t)$ and the volume airflow entering the instrument $u(t)$, which will be the three independent variables of interest in this paper. They are all represented schematically in figure \ref{schema_mouthpiece}. Several control parameters (i.e controlled by the musician) are also involved: this includes in particular the blowing mouth pressure $p_m$ and the lips' resonance frequency $\fl$.


First, the vibrating lips of the musician are described by a one-degree-of-freedom damped oscillator:

\begin{equation}
    \ddot{h} + \wlql \dot{h} + \wl^2 \pr{h-H} = \frac{p_m-p}{\mu},\\
    \label{eqh}
\end{equation}

\nin
where $\wl = 2\pi\fl$ and $\Ql$ are the angular resonance frequency and the quality factor of the lips, respectively, $\mu$ is the lips' mass per unit area, and $H$ is the lip-opening height at rest.\\

The Bernoulli's theorem applied between the mouth and the mouthpiece and taking into account the pressure drop caused by the presence of turbulence in the mouthpiece leads to the following equation:

\begin{equation}
    u(t) = wh^+(t) \sgn\pr{p_m-p(t)} \sqrt{\frac{2\md{p_m-p(t)}}{\rho}},\\
    \label{equ}
\end{equation}

\nin
with $w$ the lip-opening width (considered as constant) and $\rho$ the air density. Here, $h^+ = \max(h,0)$ accounts for the fact that the lips cannot physically interpenetrate: as soon as the lips touch ($h = 0$), the volume flow is forced to zero. The sign function $\sgn$ accounts for the possibility of air flowing from the instrument into the player's mouth.\\

Finally, the acoustic input impedance $Z(\w)$ of the resonator is described in the Fourier domain as the ratio between the pressure and the volume flow at the input of the instrument. This provides a last link between the mouth pressure and the volume airflow:

\begin{equation}
    P(\w) = Z(\w) U(\w),
    \label{eqZ}
\end{equation}

\nin
with $\w$ the angular frequency. Figure \ref{zc_AC_basse} shows the modulus and phase of $Z$ with respect to the angular frequency $\w$, for a bass trombone (A. Courtois Legend AC502).

Several limitations of this generic model have already been pointed out in previous works \cite{campbellScienceBrassInstruments2021,velutHowWellCan2017}. Firstly, it is worth noting that the instrument is described with a linear model of its input impedance: the nonlinear propagation of sound in the instrument itself is not taken into account, preventing any description of the characteristic \gu{brassy sound} of a brass instrument at high sound levels. However, since we focus here on the emergence of auto-oscillations, this effect can be neglected. Furthermore, the harmonics created by the nonlinear distortion of the wavefront propagating in the instrument are known not to be reflected back to the player when they reach the bell \cite{hirschbergShockWavesTrombones1995}. 

Secondly, lips are described as a one-degree-of-freedom damped oscillator, which is also a strong assumption. As a matter of fact, both experiments using artificial lips \cite{cullenBrassInstrumentsLinear2000} and in-vivo measurements \cite{newtonMechanicalResponseMeasurements2008} demonstrated that the lips' oscillation mechanism relies mainly on a pair of lips' resonance frequencies. Furthermore, only the first resonance of the lips reflects the \gu{outward-striking swinging door} mechanism described in \cite{campbellScienceBrassInstruments2021}, hence the single degree of freedom. In any case, the one-degree-of-freedom damped oscillator prevents any description of the \gu{buzzing phenomenon}, which corresponds to the lips' vibration without the instrument.

Thirdly, the lips' opening area $S$ in the airflow (equation \ref{equ}) is assumed to be proportional to $h$, and is thus written $S = w\times h$ with $w$ a constant. This assumption was only partially validated in \cite{bromageOpenAreasVibrating2010a}, depending on which regime is studied as well as on the sound level. In particular, a law $S \propto h^2$ would be more adequate when studying the first regimes like the pedal note.

\subsection{Numerical considerations}
\label{NT}

In this paper, we aim at determining the minimal blowing pressures for which each periodic regime (natural notes B$\flat$1, B$\flat$2, F3, B$\flat$3, D4, F4,...) is observable, as well as the corresponding playing frequencies.

In this subsection, the numerical methods allowing for the investigation of the periodic solutions of model $\br{(\ref{eqh}) \cup (\ref{equ}) \cup (\ref{eqZ})}$ are presented, as well as the modal representation of the input impedance required for the practical implementation of these methods.

\subsubsection{Numerical methods}

Several numerical methods are available to investigate the influence of a control parameter -- such as $p_m$ or $\fl$, as they are the most obvious parameters the player can change -- on the dynamics of the system. Linear stability analysis \cite{velutHowWellCan2017} consists in studying the stability of the equilibrium solutions to small perturbations, obtained by zeroing all time derivatives. This allows one to determine the threshold value of the control parameter at which the equilibrium solution destabilises, which means that at least one of the eigenvalues of the system exhibits a positive real part, and where a small-amplitude periodic regime (stable or unstable) can emerge. When a small-amplitude periodic regime indeed emerges, such a change in the system's behaviour is called a Hopf bifurcation \cite{kuznetsovElementsAppliedBifurcation2004}.

In the case of brass instruments, both $p_m$ and $\fl$ can be considered as control parameters. Figure \ref{asl_AC_basse_Hlaw} (blue curve) represents the threshold pressure predicted by the linear stability analysis with respect to $\fl$. This analysis also provides the threshold frequency, that is to say the frequency of the periodic regime that emerges when the equilibrium solution destabilises. This is represented in the bottom plot in figure \ref{asl_AC_basse_Hlaw}.

One of the main advantages of this method is its straight-forward implementation. However, it gives very little information on the oscillation regime of the model far from the so-called Hopf bifurcation point at which the equilibrium loses its stability. This method has already been used in \cite{velutHowWellCan2017}, and is applied here to the trombone (see blue curves in figure \ref{asl_AC_basse_Hlaw}).

To know more about the oscillating solution arbitrarily far from the bifurcation point (i.e the point at which the equilibrium solution becomes unstable), another approach consists in numerically solving the whole system, thanks to an ODE solver \cite{velutHowWellCan2017}. In doing so, both the transient and the stationary parts of the solution are obtained for any value of the control parameter. Nevertheless, this approach becomes tedious and unsuitable in the context of the systematic investigation of the influence of the control parameter on the oscillation regime.

Continuation methods, on the other hand, are more suitable to gain access to a more extensive view of all the oscillating regimes of the system. It consists in computing the waveform of the oscillating solution for successive values of the control parameter. The waveform corresponding to a new value of the control parameter is then deduced from previously computed waveforms, through a predictor/corrector algorithm. The behaviour of an oscillatory solution of the system with respect to a control parameter such as $p_m$ can then be assessed by plotting bifurcation diagrams, which are shown in section \ref{BDMBP}. This approach is implemented in several softwares, such as AUTO \cite{doedelAUTO97Continuation1999,gilbertMinimalBlowingPressure2020a,cochelinHighOrderPurely2009,karkarHighorderPurelyFrequency2013,freourNumericalContinuationPhysical2020} which is used in this publication. However, continuation methods require the system to be written in the form $\frac{\mathrm{d}\X}{\mathrm{d}t} = F(\X)$, with certain smoothness properties on $F$. Therefore, some work has yet to be done on the equations presented in section \ref{GBM}, which is done in the following.

\subsubsection{Input impedance}

\begin{figure}[h!]
    \centering
    \includegraphics[width=\linewidth]{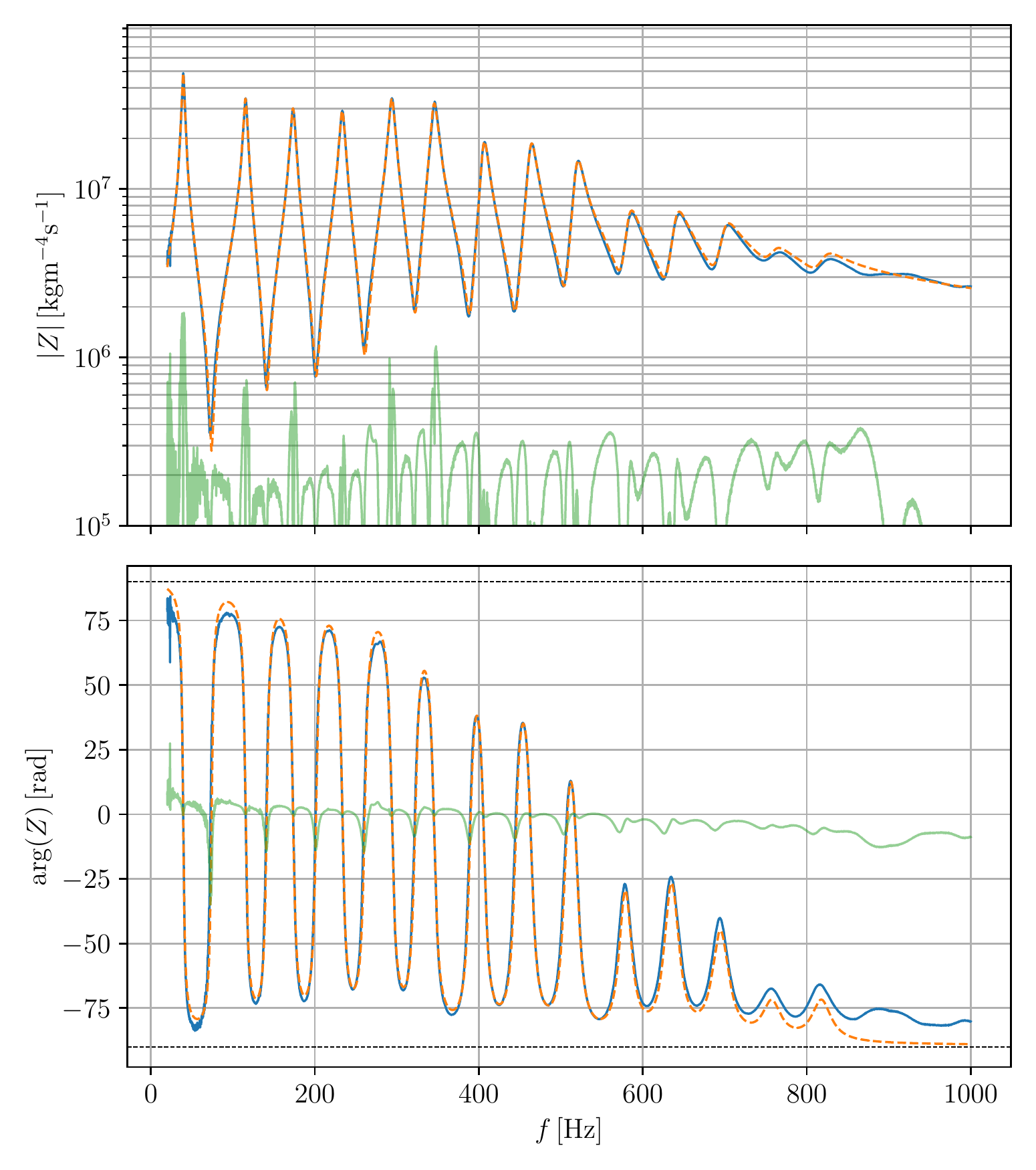}
    \caption{\baselineskip 3mm Modulus (top) and phase (bottom) of the input impedance of a bass trombone (A. Courtois Legend AC502) with respect to frequency. Blue curve: measured impedance; dashed orange curve: modal fit function with $N=15$ modes; green curve: error between measured and fitted impedances, defined by $\md{\md{\Zr}-\md{\Ze}}$ and $\arg\pr{\Zr}-\arg\pr{\Ze}$ respectively as regards the module and phase.}
    \vspace{0mm}
    \label{zc_AC_basse}
\end{figure}

Since the continuation method requires the equations to be written in the time domain, equation (\ref{eqZ}) cannot be used as it is. An analytical form of the input impedance $Z$ is therefore required to perform an inverse Fourier transform of (\ref{eqZ}). This quantity can be quite easily measured, and is represented in figure \ref{zc_AC_basse} (blue curve). This measured impedance can then be numerically fitted, in the frequency domain, by a sum of $N$ individual acoustical resonance modes of the following form \cite{ablitzerPeakpickingIdentificationTechnique2021a} :

\begin{equation}
    \Zr(\w) = \sum_{n=1}^N \frac{\J\w A_n}{\w_n^2-\w^2 + 2\J\xi_n\w_n\w}.
    \label{ZR_FA}
\end{equation}

\noindent Here, $\pr{A_n, \w_n, \xi_n} \in \R^3$ are the modal parameters of the modal decomposition. For numerical reasons, it has been chosen to convert this set of parameters $\pr{A_n,\w_n,\xi_n}$ into another one $\pr{C_n, s_n}$ so that the fitted input impedance writes ( $z^*$ is the complex conjugate of $z$):

\begin{equation}
    \Zr(\w) = Z_c \sum_{n=1}^N \pr{\frac{C_n}{\J\w-s_n} + \frac{C_n^*}{\J\w-s_n^*}},
    \label{ZC}
\end{equation}

\noindent where $Z_c$ is the characteristic input impedance of the resonator defined as $Z_c = \frac{\varrho c}{S_e}$ with $S_e$ the input cross-sectional area, and $\pr{C_n,s_n} \in \C^2$ are defined as follows:

\begin{equation}
    \left\{
        \begin{array}{r c l}
            C_n &=& \frac{A_n}{2}\pr{1 + \J\frac{\xi_n}{\sqrt{1-\xi_n^2}}},\\
            s_n &=& \w_n\pr{-\xi_n + \J\sqrt{1-\xi_n^2}}.\\
        \end{array}
    \right.
    \label{rel_CR}
\end{equation}

The new coefficients $\pr{C_n, s_n}$ also verify the relation $\Re\pr{C_n s_n^*} = 0$, since one switched from a 3-real parameters description with $\pr{A_n,\w_n,\xi_n}$ to a 2-complex parameters description with $\pr{C_n,s_n}$.

The fitted impedance $\Zr(\w)$ is plotted (orange dashed curve) in figure \ref{zc_AC_basse}. 

Reinjecting equation (\ref{ZC}) in equation (\ref{eqZ}) and applying an inverse Fourier transform (for details, see appendix \ref{IFTII}) leads to the following expression in the time domain \cite{freourNumericalContinuationPhysical2020,silvaMoReeSCFrameworkSimulation2014}:

\begin{equation}
    \dot{p}_n = Z_c C_n u + s_n p_n, \; n \in \dhk{1,N}.
    \label{eqZtemp}
\end{equation}

The mouthpiece pressure $p$ can then be written as $p = 2\sum_{n=1}^{N} \Re\pr{p_n}$. Eventually, the system $\br{(\ref{eqh}) \cup (\ref{equ}) \cup (\ref{eqZtemp})}$ is now written in the form $\frac{\mathrm{d}\X}{\mathrm{d}t} = F(\X)$, with:

\begin{equation}
    \begin{split}
        \X &= \pr{\br{X_m}_{m \in \dhk{1,2(N+1)}}}\\
           &= \pr{ h; \dot{h}; \br{\Re(p_n)}_{n \in \dhk{1,N}}; \br{\Im(p_n)}_{n \in \dhk{1,N}}}
    \end{split}
\end{equation}

\nin
the state vector, so that $h = X_1$, $\hp = X_2$, and $p = 2\sum_{n=1}^{N} \Re\pr{p_n} = 2\sum_{n=3}^{N+2}X_n$. Taking the real and imaginary part of the $N$ equations (\ref{eqZtemp}) yields $2N$ real equations, so that the nonlinear function $F$ is defined as:

\begin{equation}
    F:\X\mapsto
    \begin{pmatrix}
        X_2\\
        - \gaml X_2 - \wl^2\pr{X_1-H} + \frac{p_m-2\sum_{n=3}^{N+2}X_n}{\mu}\\
        \Re\hk{s_1\pr{X_3 + \J X_{N+3}} + Z_c C_1 u(\X)}\\
        \Re\hk{s_2\pr{X_4 + \J X_{N+4}} + Z_c C_2 u(\X)}\\
        \vdots\\
        \Re\hk{s_N\pr{X_{N+2} + \J X_{2(N+1)}} + Z_c C_N u(\X)}\\
        \Im\hk{s_1\pr{X_3 + \J X_{N+3}} + Z_c C_1 u(\X)}\\
        \Im\hk{s_2\pr{X_4 + \J X_{N+4}} + Z_c C_2 u(\X)}\\
        \vdots\\
        \Im\hk{s_N\pr{X_{N+2} + \J X_{2(N+1)}} + Z_c C_N u(\X)}\\
    \end{pmatrix},
    \label{sys}
\end{equation}

with $u:\X \mapsto w X_1^+ \sgn\pr{p_m-2\sum_{n=3}^{N+2}X_n} \times$ $\sqrt{\frac{2}{\varrho}\md{p_m-2\sum_{n=3}^{N+2}X_n}}$.

\subsubsection{Regularisation of the volume airflow}

For the continuation method implemented in AUTO to work correctly, $F$ has to be a $\mathcal{C}^1$ vector function. In this respect, noting that $\sgn(x) = x/\md{x}$ for $x \ne 0$ and $x^+ = \max\pr{x,0} = \frac{1}{2}\pr{x+\md{x}}$, equation (\ref{equ}) is regularised based on the regularisation $\md{x} \underset{\eta \to 0}{\sim} = \sqrt{x^2+\eta}$ used in section 2 of \cite{colinotInfluenceGhostReed2019}:

\begin{equation}
    u \underset{\eta \to 0}{\sim} w \times \frac{h + h_0\sqrt{\pr{\frac{h}{h_0}}^2+\eta}}{2} \times \frac{p_m-p}{\sqrt{p_0}\sqrt[4]{\pr{\frac{p_m-p}{p_0}}^2+\eta}},
    \label{equreg}
\end{equation}

or equivalently

\begin{equation}
    \begin{split}
        u &\underset{\eta \to 0}{\sim} w \times \frac{X_1 + h_0\sqrt{\pr{\frac{X_1}{h_0}}^2+\eta}}{2}\\
          &\hspace{5mm}\times \frac{p_m-2\sum_{n=3}^{N+2}X_n}{\sqrt{p_0}\sqrt[4]{\pr{\frac{p_m-2\sum_{n=3}^{N+2}X_n}{p_0}}^2+\eta}}
    \end{split}
    \label{equregX}
\end{equation}

\noindent
in terms of the components of the state vector $\X$, with $\eta$ the regularisation parameter, which is fixed to \num{e-6} in the following. $p_0$ is defined in a similar way as the closure pressure for woodwind instruments \cite{gilbertMinimalBlowingPressure2020a}: $p_0 = \mu \w_0^2 h_0$, except that $\w_0$ is chosen close to the first resonance frequency of the resonator. Indeed, the choice of $\wl$ generally considered for woodwind instruments is not suitable in the case of brass instruments where $\wl$ is no longer constant. In practice, the choice is not $\w_0 = \w_{\text{res},1}$ (where $\br{\w_{\text{res},n}}_{n\in\dhk{1,N}}$ are the resonance angular frequencies of the resonator), but rather $\w_0 = \w_{\text{res,4}}/4$, because the fourth resonance of instruments having the same tube length (a trombone and a euphonium for instance) appears to be quite constant, contrary to the first resonance frequency which can vary up to 8 semitones between a trombone and euphonium.

In the following, the system equations (\ref{sys}) with $u(\X)$ given by equation (\ref{equregX}) are processed numerically in a dimensionless form, which is detailed in appendix \ref{DVEUPP}. Typical bifurcation diagrams will be shown in section \ref{BDMBP}.

\section{Lips' parameters and linear stability analysis}
\label{LPLSA}

The results of the linear stability analysis detailed in subsection \ref{CLOHR} suggest that the lips' parameter $\fl$ can be tuned so as to obtain a given desired oscillatory regime.
The values of the other lips' parameters are difficult to measure and thus hardly found in the literature. In a first subsection, we consider the same parameters as in \cite{velutHowWellCan2017}, which are detailed in table \ref{lips_params}. In order to be more consistent with experimental results, we introduce, in a second subsection; a frequency-dependent lip-opening at rest, based on experimental data from \cite{elliottRegenerationBrassWind1982}.

\subsection{Constant lip-opening height at rest}
\label{CLOHR}

In the case of constant lips' parameters, the results of the linear stability analysis of the system are represented in figure \ref{asl_AC_basse_Hlaw} by the blue U-shaped patterns. As described in subsection \ref{NT}, these results give the values of the mouth pressure $p_m$ at which the equilibrium solution destabilises (the threshold mouth pressures), as well as the related frequency at which the oscillating solution emerges (the threshold frequencies). As a matter of fact, the multiple U-shaped patterns on the top plot of figure \ref{asl_AC_basse_Hlaw} reflect the fact that for a given configuration of the resonator (i.e without moving the slide nor activating any trigger), the trombone player can play several notes called \gu{natural notes} just by changing the lips' resonance frequency (see for instance \cite{campbellScienceBrassInstruments2021}) represented in figure \ref{partiels}, from the B$\flat$1 (\SI{58}{\hertz}) to B$\flat$4 (\SI{466}{\hertz}) approximately. For each natural note or regime $n$ to emerge, there is an optimal lips' resonance frequency $\flo{eq}$ which corresponds to the minimum of a U-shaped pattern. This is associated with an optimal threshold mouth pressure $\pso{eq}$ and an optimal threshold frequency $\fso{eq}$. The notation $\flo{eq}$ has been chosen to be as consistent as possible with \cite{velutHowWellCan2017}: in this article, the equilibrium optimal threshold value (i.e.  the threshold inferred from the linear stability analysis) of a quantity $q$ in the $n^{\text{th}}$ regime was written $q_{\text{thresh},n}^{\text{opt}}$. Here, the threshold quantities are obtained either using linear stability analysis or through a bifurcation analysis, which will be defined and addressed in section \ref{BDMBP}. Therefore, it has been chosen to add an extra superscript \gu{eq} for \gu{equilibrium}, referring to the linear stability analysis, or \gu{per} for \gu{periodic solutions}, referring to the analysis of bifurcation diagrams. Furthermore, it can be noticed that the patterns on the bottom plot of figure \ref{asl_AC_basse_Hlaw} are always above the line whose equation is $\fs = \fl$, which is a characteristic of the outward-striking valve model: the instrument always plays a note slightly above the lips' resonance frequency.


\subsection{Frequency-dependent lip-opening height at rest}

\begin{figure}[h!]
    \centering
    \includegraphics[width=\linewidth]{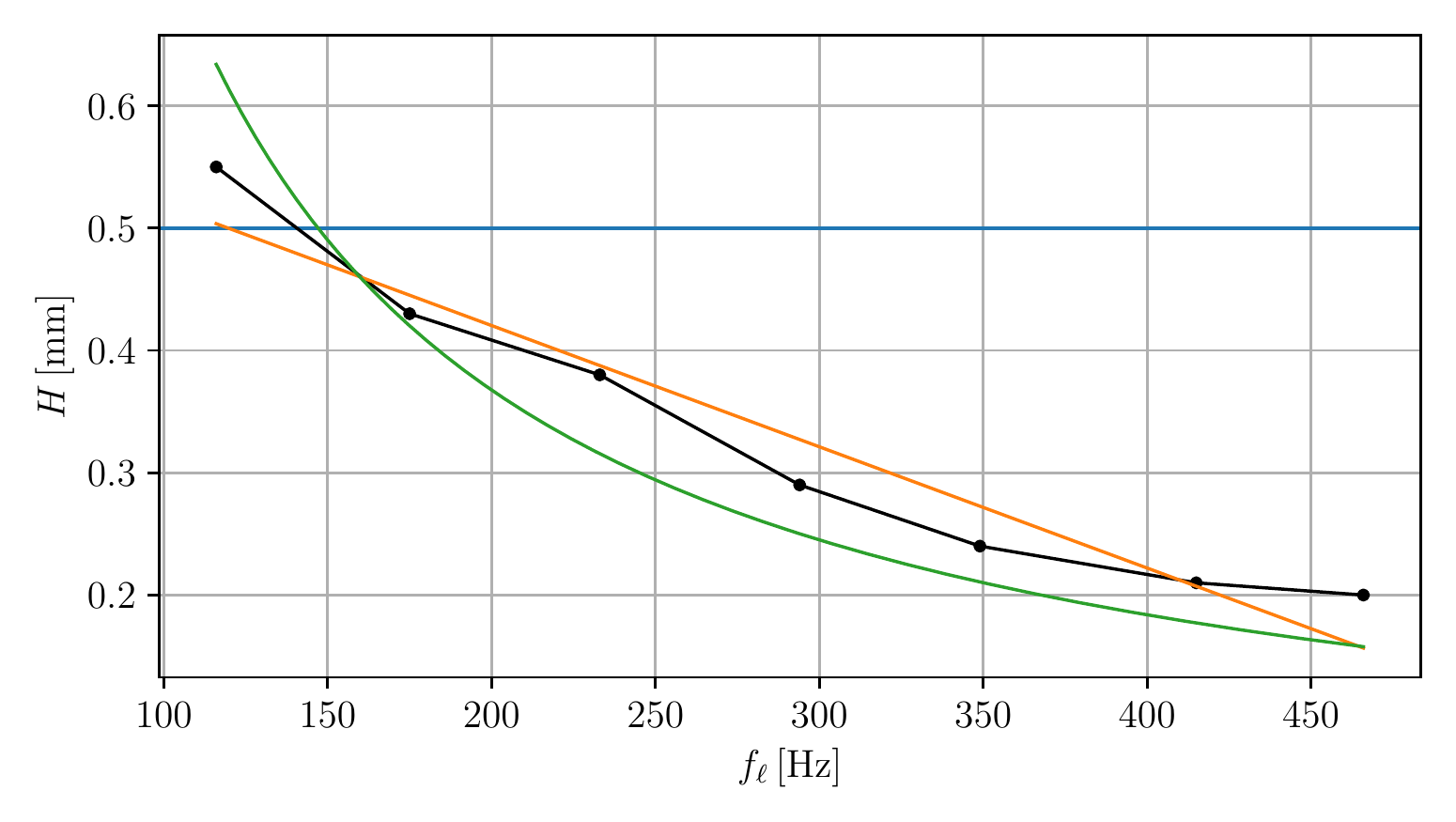}\\
    \caption{\baselineskip 3mm Considered $H$-profiles vs. lips' resonance frequency. Blue line corresponds to the constant value of $H = h_0$ from \cite{velutHowWellCan2017}, black points correspond to experimental data from \cite{elliottRegenerationBrassWind1982}, and the orange and green curves correspond to the regression functions chosen to fit the experimental points, respectively $H = a\fl + b$ and $H \propto 1/\fl$.}
    \vspace{0mm}
    \label{H_fits_EB}
\end{figure}

\begin{figure}[h!]
    \centering
    \includegraphics[width=\linewidth]{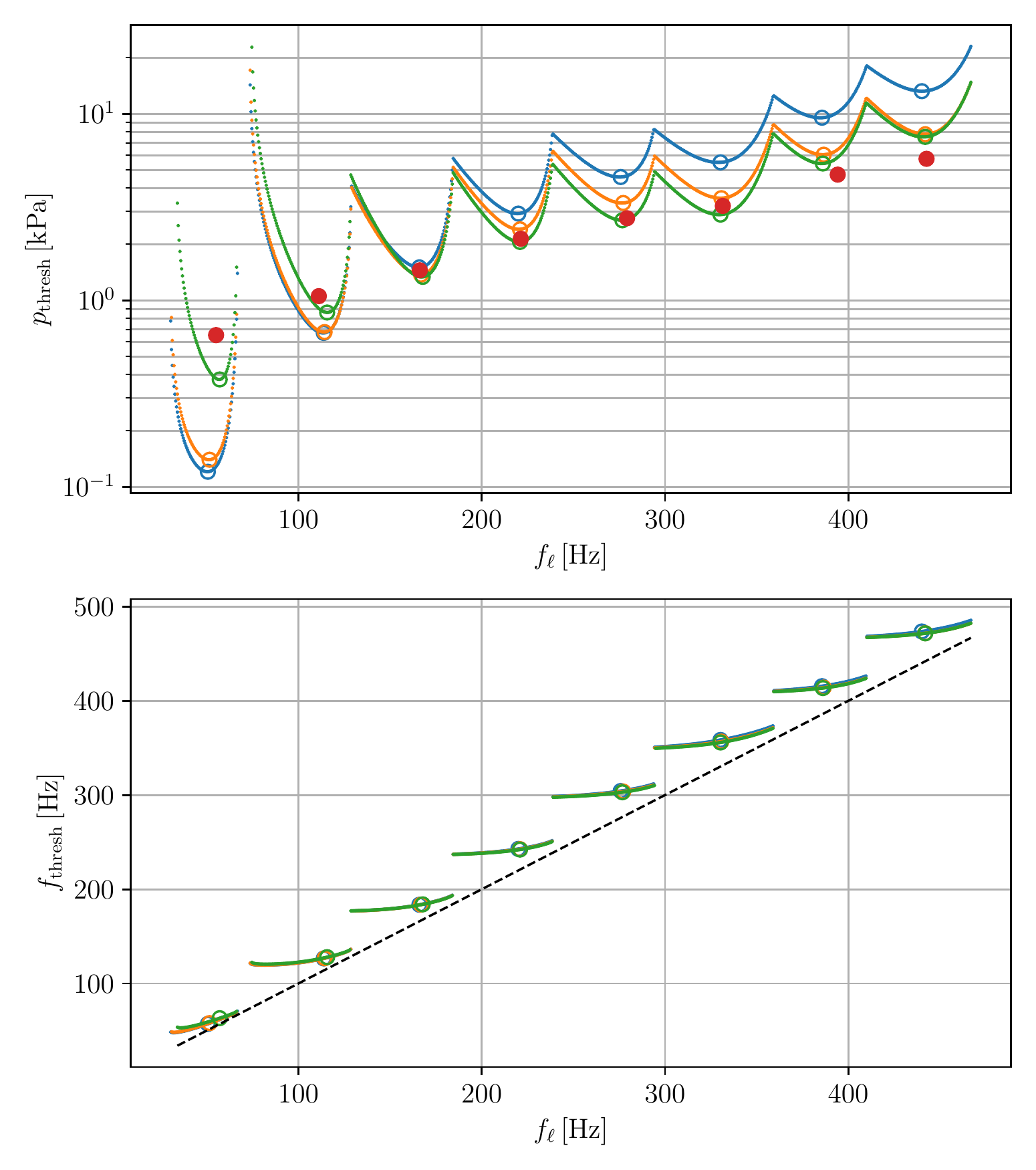}
    \caption{\baselineskip 3mm Results of the linear stability analysis with the constant, linear and inverse $H$-profile described on figure \ref{H_fits_EB}. Top and bottom plots represent respectively the threshold mouth pressure and the threshold frequency vs. the lips' resonance frequency. Blue: $H = h_0$, orange: $H = a\fl + b$, green: $H \propto 1/\fl$. Circles point out the minima of each U-shaped pattern on the top plot. Red points represent the experimental values of minimal threshold pressures given by \cite{gilbertAnalyseStabiliteLineaire2018}. The black dashed line on the bottom plot represents $\fs = \fl$.}
    \vspace{0mm}
    \label{asl_AC_basse_Hlaw}
\end{figure}

\begin{figure}[h!]
    \centering
    \includegraphics[width=0.7\linewidth]{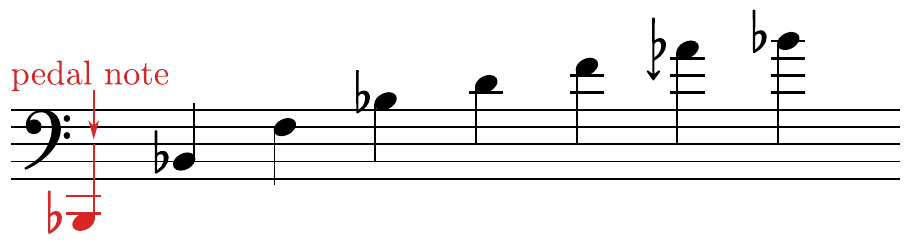}\includegraphics[width=0.3\linewidth]{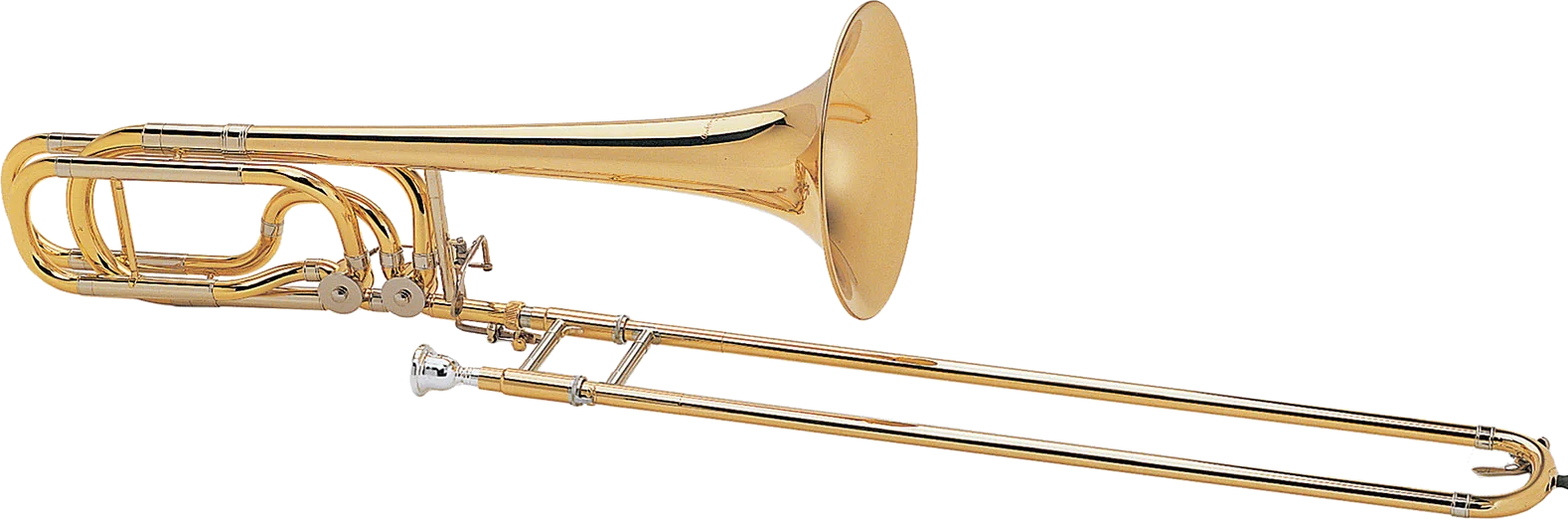}
    \caption{\baselineskip 3mm Natural notes playable by a trombone, i.e without moving the slide nor activating any trigger. From left to right: B$\flat$1 (pedal note), B$\flat$2, F3, B$\flat$3, D4, F4, A$\flat$4 (naturally a bit flat compared to an equal tempered scale), B$\flat$4.}
    \label{partiels}
\end{figure}

In order to assess the validity of the results presented above, a comparison is drawn with the minimal threshold mouth pressures estimated experimentally in \cite{gilbertAnalyseStabiliteLineaire2018}, for the exact same trombone as the one considered in this paper. In figure \ref{asl_AC_basse_Hlaw}, the comparison between the minima of the blue U-shaped patterns and the red dots (representing the experimental measures of the minimal threshold mouth pressures) highlights significant differences. In particular, the results of the linear stability analysis, on the one hand, and of the experiment, on the other hand, are quite different: compared to the experimental data, threshold pressures computed numerically are too low for the first regimes, and too high for the last regimes. To overcome this issue, additional information coming from experiments in \cite{elliottRegenerationBrassWind1982} is considered. Indeed, the lips' opening height at rest $H$ is measured to decrease monotonously with $\fl$. The experimental points are plotted in black on the top plot of figure \ref{H_fits_EB}, together with two regression functions which aim at fitting these points.
The inverse function $H = k/\fl$ ($k$ is given in table \ref{lips_params}) appears here to better fit the experimental points than the linear regression $H = a\fl + b$. Therefore, the former $H$-profile will be retained hereafter, even if the case $H = h_0$ is still given alongside for sake of comparison until the end of section \ref{BDMBP}.

\begin{table}[h!]
    \centering
    \begin{tabular}{|c|c|c|c|}
        \hline
        $h_0\,[\si{\metre}]$ & $w\,[\si{\metre}]$ & $\Ql$ & $\frac{1}{\mu}\,[\si{\square\metre\per\kilogram}]$\\
        \hline
        \num{5,0e-4} & \num{1,2e-2} & \num{7,0} & \num{1,1e-1} \\
        \hline
        \hline
        $\varrho\,[\si{\kilogram\per\cubic\metre}]$ & $c_0\,[\si{\metre\per\second}]$ & $k\,[\si{\metre\hertz}]$ &\\
        \hline
        \num{1,2} & \num{3.4e2} & \num{7.4e-2} &\\
        \hline
    \end{tabular}
    \caption{\baselineskip 3mm Lips' parameters used in the present publication, mostly taken from \cite{velutHowWellCan2017}.}
    \label{lips_params}
\end{table}

\section{Bifurcation diagrams and minimal blowing pressure}
\label{BDMBP}

In this section several typical bifurcation diagrams of a trombone are shown and discussed in detail. The additional information obtained with the continuation method on the threshold mouth pressure is also discussed, leading to new conclusions compared to the case of the linear stability analysis.

\subsection{Case of the third regime}
\label{BDTR}

In this section the example of the third regime of a trombone is considered (F3, third U-shaped pattern on top plot of figure \ref{asl_AC_basse_Hlaw}), as it shows a variety of different behaviours of the system.

\begin{figure*}[h!]
    \centering
    \includegraphics[width=0.5\linewidth]{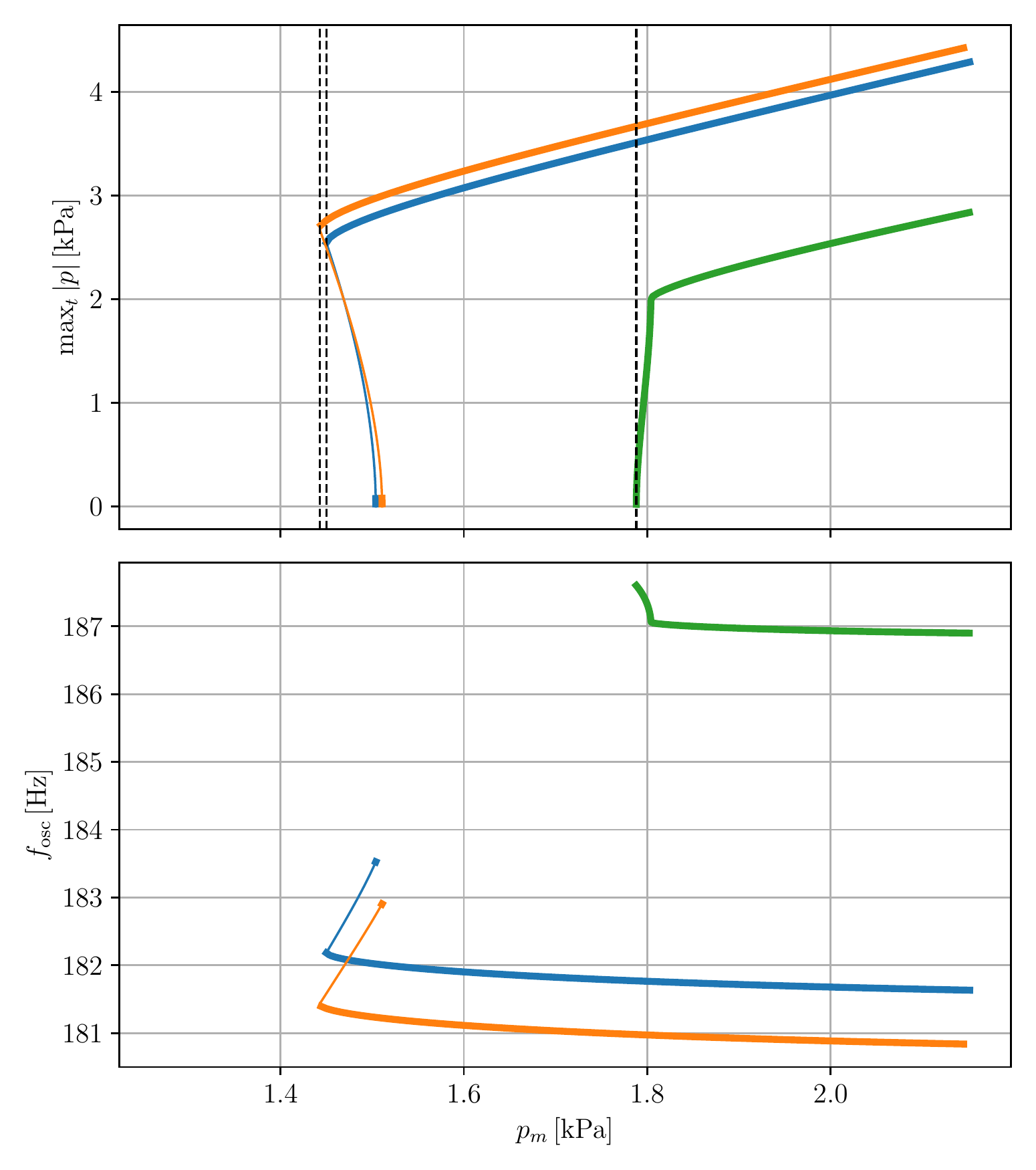}\includegraphics[width=0.5\linewidth]{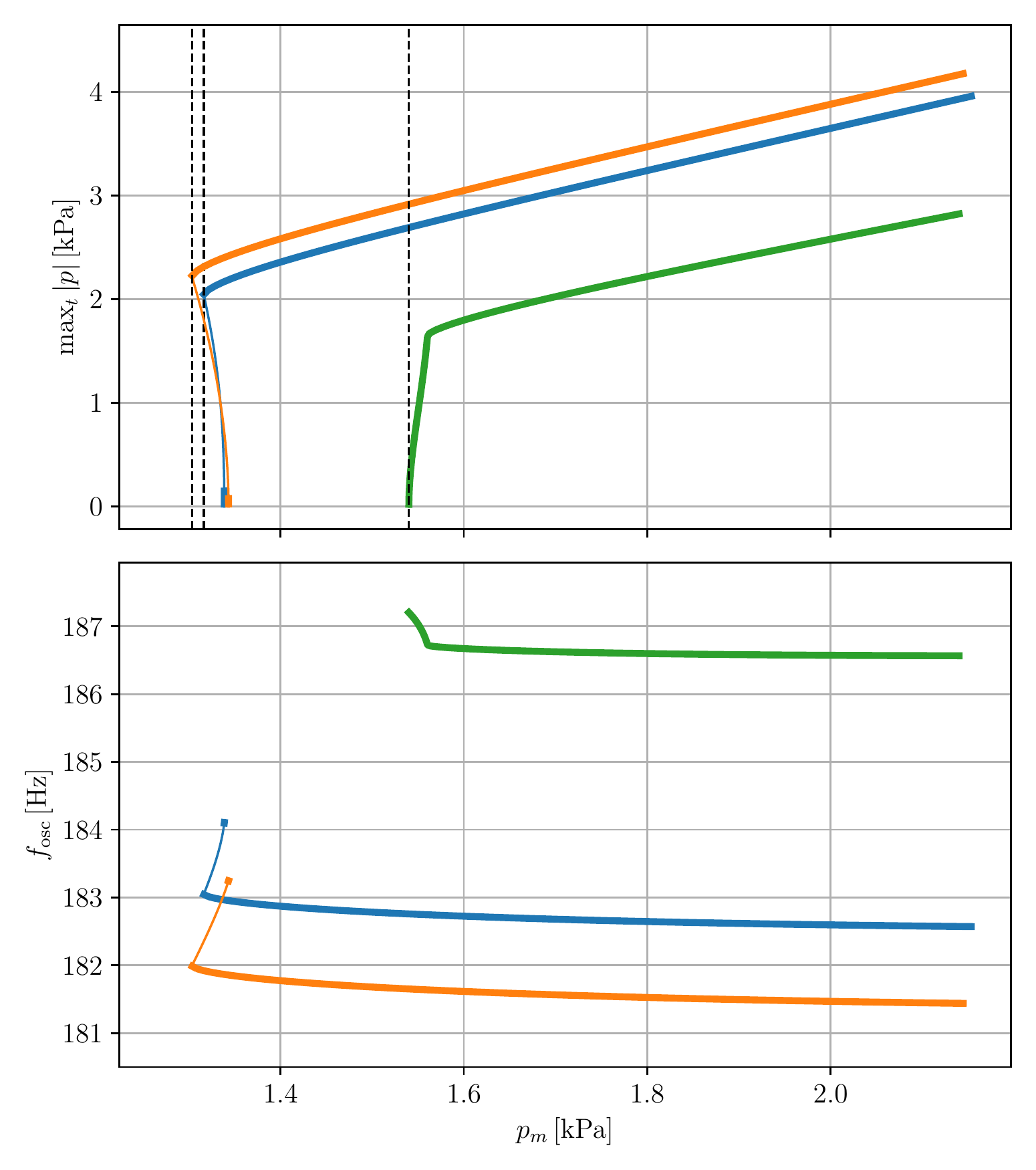}
    \caption{\baselineskip 3mm Top and bottom plots represent respectively the maximum amplitude of the periodic oscillation branches and their oscillation frequency vs. the blowing pressure. Left: case $H = h_0$; right: case $H \propto 1/\fl$. Blue curves: case $\fl = \flon{eq}{3} = \SI{165.84}{\hertz}$ (left) or $\SI{167.75}{\hertz}$ (right); orange curves: case $\fl = \flon{per}{3} = \SI{164.00}{\hertz}$ (left) or $\SI{165.17}{\hertz}$ (right); green curves: case $\fl = \SI{175.09}{\hertz}$, which corresponds to an arbitrary $\fl$ value at which the Hopf bifurcations are direct. The line thickness indicates whether the branch portion is stable (thick line) or unstable (thin line). The vertical dashed lines locate the value of the minimal threshold mouth pressure of each curve.}
    \vspace{0mm}
    \label{bd_AC_basse}
\end{figure*}

The left plot of figure \ref{bd_AC_basse} shows the bifurcation diagrams obtained by continuation for a constant $H$-profile $H = h_0$ and for three different but close values of $\fl$. For each branch of periodic solution, there is a global minimum of the mouth pressure (identified by the vertical dotted lines) below which the regime does not exist anymore. This minimum can be lower than or equal to the linear threshold mouth pressure given by the linear stability analysis in section \ref{LPLSA}. As for this latter method, an optimal threshold mouth pressure $\pso{per}$ \footnote{see subsection \ref{CLOHR} for details about this notation.} which can be lower than or equal to $\pso{eq}$ is introduced, as well as the associated optimal threshold frequency $\fso{per}$ and the associated optimal lips' resonance frequency $\flo{per}$. For instance, in the case $\fl = \SI{175.09}{\hertz}$ (green curve) the observed bifurcation is a direct Hopf \cite{kuznetsovElementsAppliedBifurcation2004} and the minimal blowing pressure is the same as the one given by the linear stability analysis. On the contrary, in the case $\fl = \flon{eq}{3}$ (blue curve) and $\fl = \flon{per}{3}$ (orange curve), a small stable portion at the very beginning of the branch is noticed, thus rigorously classifying the bifurcation as a direct Hopf. However, since this stable portion only exists for a narrow range of $p_m$ it would be very difficult to observe it experimentally, hence we choose to neglect it. In the following, this configuration will be referred to as \gu{almost-inverse Hopf} instead. In the case of this almost-inverse Hopf bifurcation, the solution emerging from the equilibrium is unstable (if we neglect the small stable portion at the very beginning) and stabilises further on the branch at a value of threshold mouth pressure lower than at the equilibrium. For $\fl = \flon{per}{3}$, the actual minimal blowing pressure is not only below the linear threshold pressure but also below the minimal blowing pressure given by the case $\fl = \flon{eq}{3}$.

However, it is worth noting that the linear stability analysis provides a rather accurate first guess of the note which is the easiest to play. Indeed, the frequency interval between the easiest note according to the linear stability analysis (corresponding to the minimal blowing pressure of the blue curve on figure \ref{bd_AC_basse}) and according to the bifurcation diagrams (corresponding to the minimal blowing pressure of the orange curve on figure \ref{bd_AC_basse}) is less than a semitone.

Eventually, it is reassuring to note that even if the variable $H$-profile $H \propto 1/\fl$ has a significant impact on the threshold pressure of the system, it has nonetheless qualitatively no influence on its threshold frequencies as shown on bottom plot of figure \ref{asl_AC_basse_Hlaw}.

For the interested reader, an animation showing the evolution of the bifurcation diagrams as a function of the lips' resonance frequency $\fl$ is provided online\footnote{Animation showing the evolution of the bifurcation diagrams as a function of the lips' resonance frequency $\fl$ increasing from \SI{134.2}{\hertz} to \SI{182.1}{\hertz} in the case of the constant $H$-profile $H=h_0$ at \url{http://perso.univ-lemans.fr/~rmatte/bd_dyn_AC_basse_reg_reg3.avi} or in the case of the inverse $H$-profile $H \propto 1/\fl$ at \url{http://perso.univ-lemans.fr/~rmatte/bd_dyn_AC_basse_Hinv_reg_reg3.avi}.}. It especially shows how the bifurcation can go from almost-inverse Hopf to direct Hopf through the merging of the two stable portions of the branch into a single one.

\subsection{Minimal blowing pressures}
\label{MBP}

\begin{figure*}[h!]
    \centering
    \includegraphics[width=0.5\linewidth]{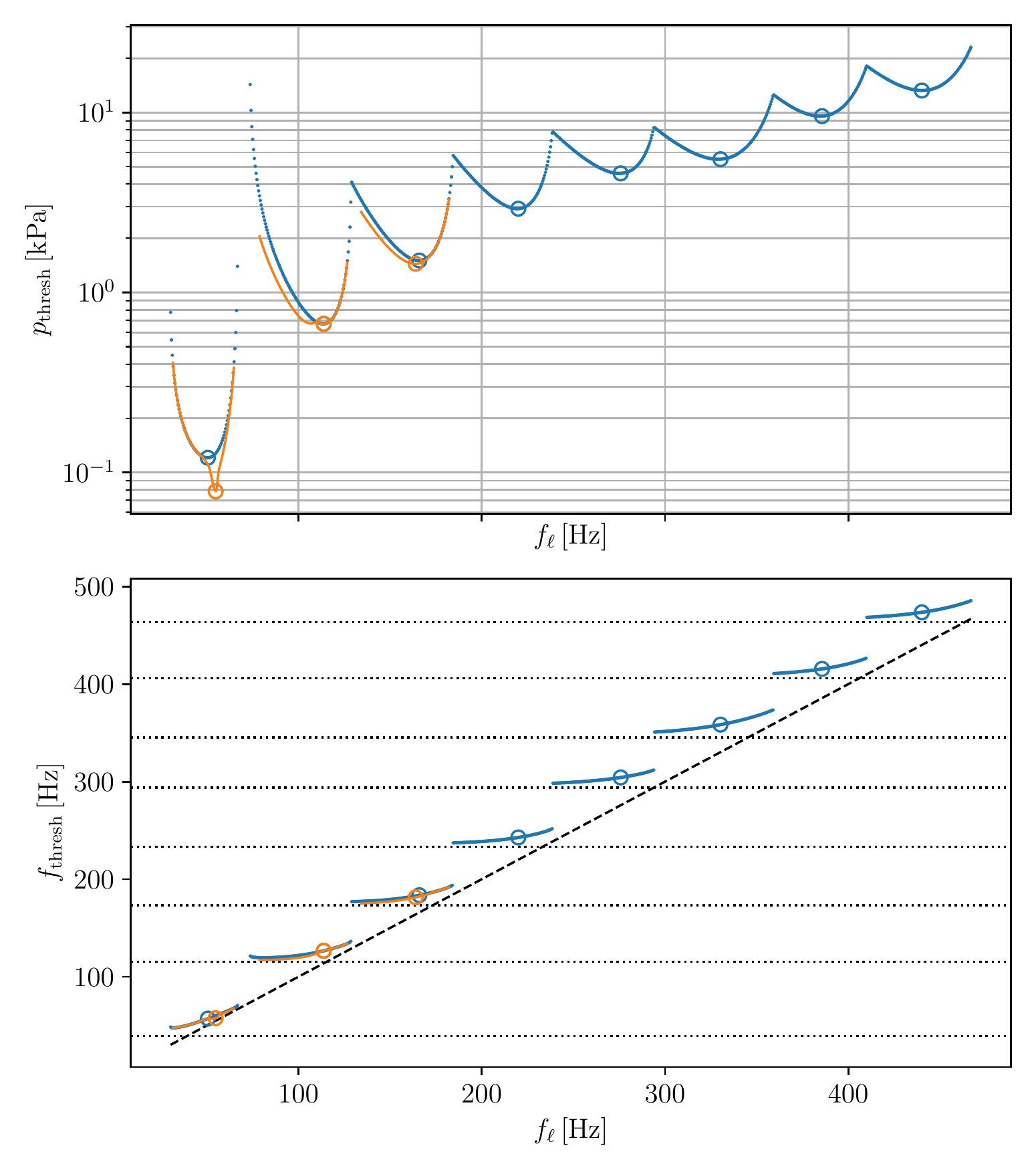}\includegraphics[width=0.5\linewidth]{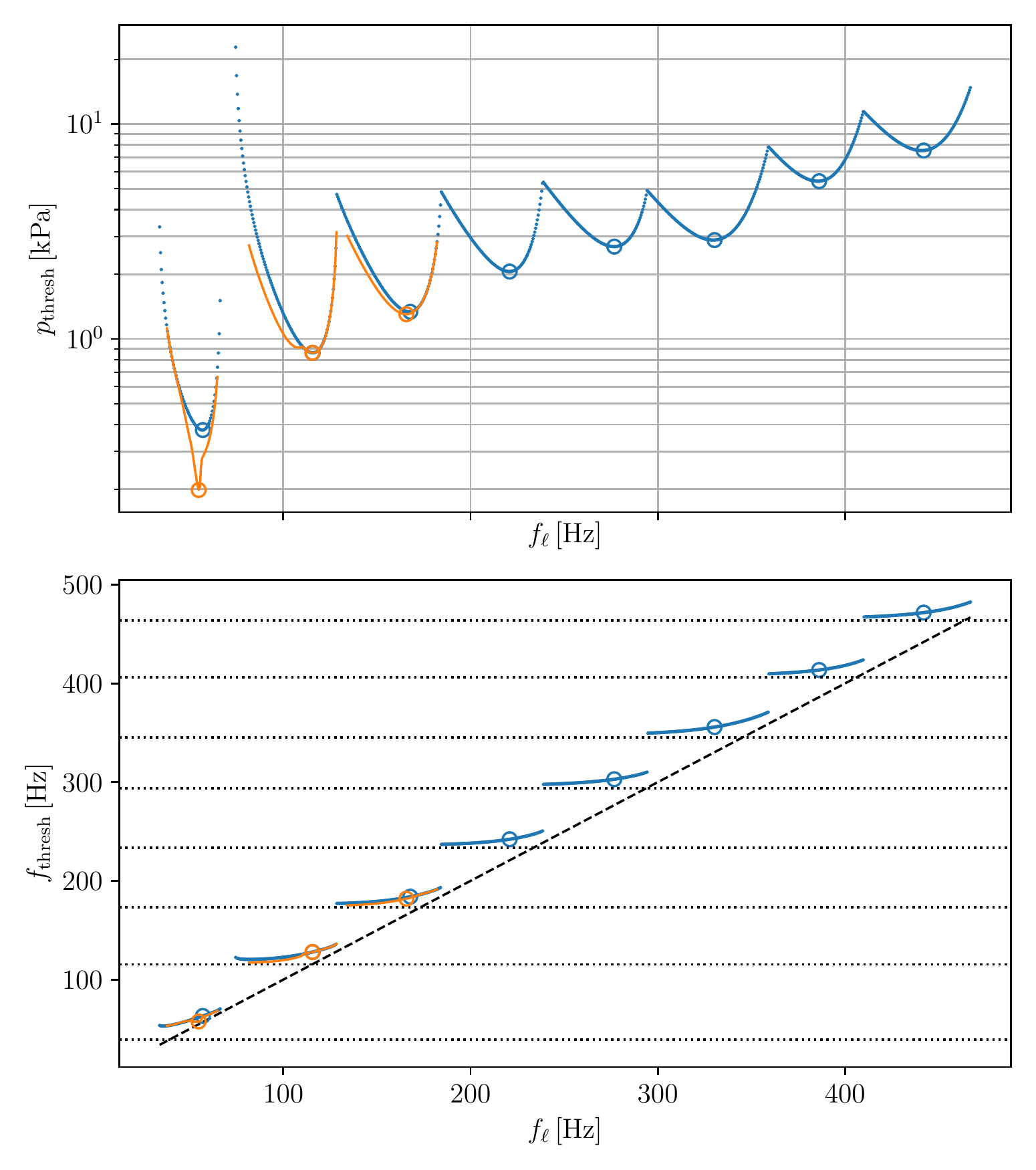}
    \caption{\baselineskip 3mm Top and bottom plots represent respectively the threshold pressures and threshold frequencies given by the linear stability analysis (blue) and the continuation method (orange) vs. the lips' frequency. Left: case $H = h_0$; right: case $H \propto 1/\fl$. The circles identify the value of $\fl$ corresponding to a local minimum of blowing pressure. Horizontal dotted lines on the bottom plots locate the values of the acoustical resonances of the resonator; the black dashed line on the bottom plots represents $\fs = \fl$.}
    \vspace{0mm}
    \label{psc_vs_asl_AC_basse}
\end{figure*}

The investigation performed in subsection \ref{BDTR} for two values of $\fl$ is repeated for a wide range of lips' resonance frequency. For each value of $\fl$, the global minimal blowing pressure is extracted from the bifurcation diagrams, and these threshold values are compared with the results  of linear stability analysis.
The results of both methods are represented in figure \ref{psc_vs_asl_AC_basse}, which clearly highlights the added value of the continuation approach. In particular, the fact the first regimes (B$\flat$1, B$\flat$2 and F3) emerge through inverse Hopf bifurcations results in an actual threshold pressure which is lower than the one predicted by the linear stability analysis. Nevertheless, it is worth noting that the system only exhibit direct Hopf bifurcations for regimes $\ge 4$. This means that the linear stability analysis is sufficient to characterise the threshold mouth pressures for these regimes.
Nevertheless, it is worth noting that similar investigations performed for a trumpet (which is very similar to a trombone considering its bore, except that its pitch is an octave higher) in \cite{freourNumericalContinuationPhysical2020} highlighted an inverse Hopf bifurcation for the fourth regime. Also, the behaviour observed here for the trombone has also been observed in the case of the euphonium or the saxhorn, except for the first regime which will be addressed in section \ref{FR}.

As in subsection \ref{BDTR}, it is reassuring here to note that the variable $H$-profile $H \propto 1/\fl$ has qualitatively no influence on the system's behaviour discussed above, in particular its threshold frequencies as shown on bottom plots of figure \ref{psc_vs_asl_AC_basse}.

Among all the regimes that can be played by the musician according to figure \ref{psc_vs_asl_AC_basse} and figure \ref{partiels}, the first one is especially worth focusing on. Indeed, it exhibits a very different behaviour depending on whether the instrument has a predominantly-cylindrical bore profile (case of the trombone or the trumpet for instance) or a predominantly-expanding bore profile (case of the tubas or the flugelhorn); see for instance figure 7.34 of \cite{campbellScienceBrassInstruments2021}. In this respect, the last section of this paper focuses on the study of the first regime of the trombone (predominantly-cylindrical bore profile) and the euphonium (predominantly-expanding bore profile).

\section{Bifurcation diagrams, case of the first regime}
\label{FR}

\subsection{Case of the trombone}

\begin{figure}[h!]
    \centering
    \includegraphics[width=\linewidth]{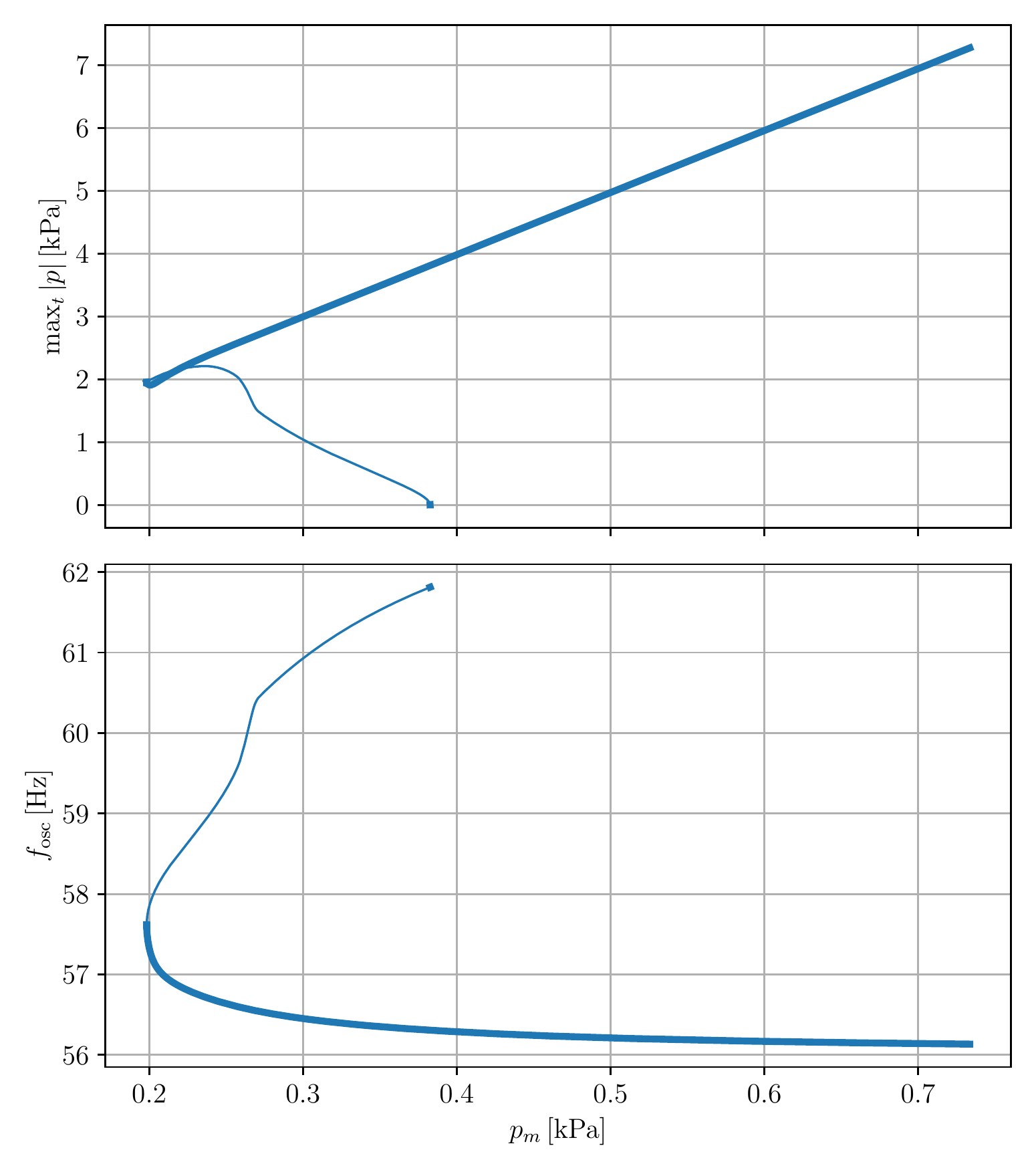}
    \caption{\baselineskip 3mm Top and bottom plots represent respectively the maximum amplitude of the periodic solutions branches and their oscillation frequency vs. the blowing pressure in the case of a trombone, for $\fl = \flon{per}{1} = \SI{54.86}{\hertz}$. The line thickness indicates whether the branch portion is stable (thick line) or unstable (thin line). }
    \vspace{0mm}
    \label{bd_AC_basse_reg1}
\end{figure}

Similarly to the case of the third regime (see subsection \ref{BDTR}), the bifurcation diagrams of the trombone show that the  first regime generally emerges through an inverse Hopf bifurcations (see figure \ref{bd_AC_basse_reg1}). As already discussed in subsection \ref{MBP}, this leads to an optimal threshold pressure $\pson{per}{1}$ smaller than $\pson{eq}{1}$. However, it is worth noting that, similarly to other regimes, only one note is accessible to the musician since only one branch of the bifurcation diagrams is stable. The corresponding note is called the \gu{pedal note} (B$\flat$1, lowest note playable in first position and roughly one octave below the second regime B$\flat$2, see also figure \ref{partiels}).

The optimal playing frequency $\fson{per}{1}$ of the pedal note predicted by the bifurcation diagrams is about a semitone lower than the optimal linear threshold frequency $\fson{eq}{1}$ given by the linear stability analysis used in \cite{velutHowWellCan2017}. Therefore, the analysis of the bifurcation diagrams adds a significant information as regards the playing frequency, even though -- as previously discussed in subsection \ref{MBP} -- the linear stability analysis already provides a rather good estimation of the playing frequency.

\subsection{Case of a bass brass instruments with predominantly-expanding bore profile}

For the regimes 2 to 8, the bifurcation diagrams of the euphonium (not shown here) are very similar to the trombone's ones, in terms of the direct or inverse nature of the Hopf bifurcations. However, the euphonium exhibits a specific behaviour for the first regime, which is qualitatively different from the one observed for the trombone.

\begin{figure}[h!]
    \centering
    \includegraphics[width=0.7\linewidth]{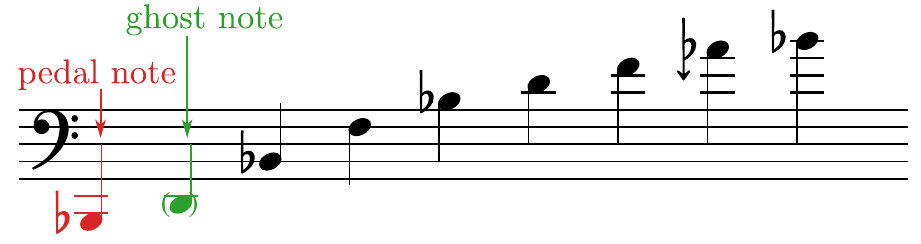}\includegraphics[width=0.3\linewidth]{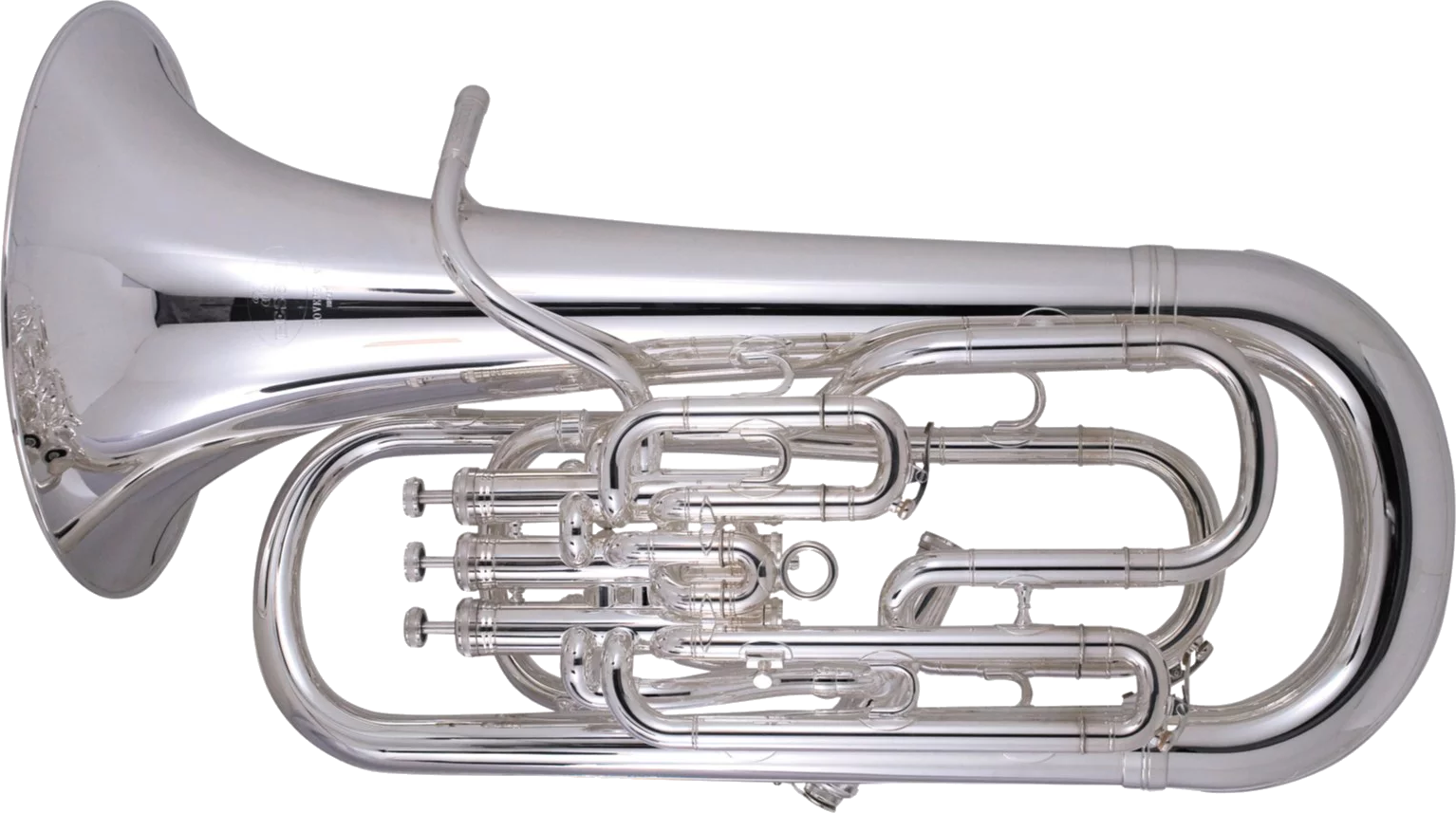}
    \caption{\baselineskip 3mm Natural notes playable by a euphonium, i.e without depressing any valves. From left to right: B$\flat$1 (pedal note, red), D2 (ghost note, green), B$\flat$2, F3, B$\flat$3, D4, F4, A$\flat$4 (naturally a bit flat compared to an equal tempered scale), B$\flat$4.}
    \label{partiels_gn}
\end{figure}

\begin{figure*}[h!]
    \centering
    \includegraphics[width=\linewidth]{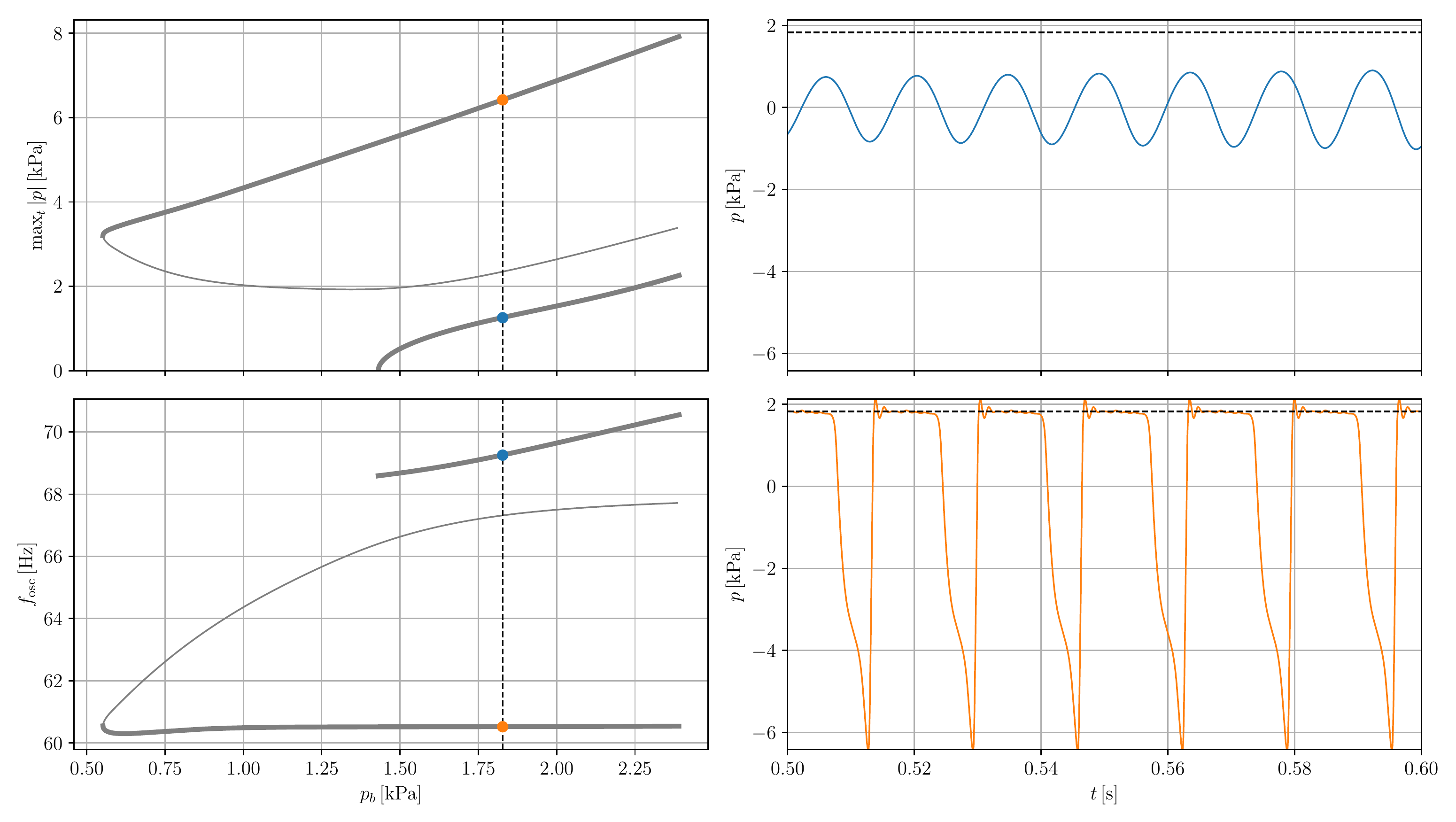}
    \caption{\baselineskip 3mm Typical bifurcation diagrams (left) and time-domain simulations (right) in the case of the euphonium's first regime, obtained for $p_m = \SI{1.8}{\kilo\pascal}$ and $\fl = \flon{per}{\text{PN}} (\text{Pedal Note}) = \SI{49}{\hertz}$. Left plots represent the maximum amplitude of the periodic oscillation branches vs. the blowing pressure (top) and the frequency of the corresponding periodic solutions vs. the blowing pressure (bottom). The line thickness indicates whether the branch portion is stable (thick line) or unstable (thin line). Right plots represent the time evolution of the acoustic pressure at two points of the bifurcation diagrams above, given by the intersection of the vertical dotted line and the branches.  Top: time-domain simulation of the ghost note (blue point on the bifurcation diagrams); bottom: time-domain simulation of the pedal note (orange point on the bifurcation diagrams). The horizontal dotted line represents the value of $p_m$.}
    \label{simutemp_nfp}
\end{figure*}

In the case of the euphonium, the left part of figure \ref{simutemp_nfp} shows typical bifurcation diagrams, obtained for a value of lips' resonance frequency $\fl$ in the range of the first regime. Contrary to the trombone, these diagrams display two stable regimes corresponding to two distinct periodic solution branches, thus resulting in two different playing regimes accessible to the musician. The note with the lowest oscillation frequency is called -- as for the trombone -- the \gu{pedal note} (B$\flat$1, lowest note playable without any valve pushed, see also figure \ref{partiels_gn}), whereas the other one is referred to as the \gu{ghost note} \footnote{We chose here to use the term employed by \cite{velutHowWellCan2017} to qualify this note being between a minor third and a perfect fourth higher than the pedal note.} (see figure \ref{partiels_gn}).

On the one hand, the pedal note (red note on figure \ref{partiels_gn}) is a natural note, which is as familiar as any other regime for both trombone and euphonium players. On the other hand, the ghost note (green note on figure \ref{partiels_gn}) is a regime exclusively accessible to bass brass instruments with predominantly-expanding bore profile as a whole, such as the euphonium. Even though the ghost note is not much used in a musical context by euphonium players, it can be played by any tuba player and was brought to light for the first time in \cite{velutHowWellCan2017} using linear stability analysis. In particular, it was the only note playable for values of $\fl$ close to the first acoustic resonance of the resonator according to this method. As a matter of fact, since the branch corresponding to the pedal note (see the left plots of figure \ref{simutemp_nfp}) does not emerge from the equilibrium solution (whereas the branch of the ghost note does), this regime could not be found by linear stability analysis which gives trustworthy information only in the vicinity of an equilibrium.
Furthermore, the ghost note can be linked to one of the \gu{factitious notes} mentioned in \cite{herbertCambridgeEncyclopediaBrass2019}, where it is referred to as a \gu{phenomenon whereby a player can sound a note intermediate between notes low in the series of natural notes} and sounds a perfect fourth above the pedal note in the case of an E$\flat$-tuba.

To illustrate the uncommon behaviour of the bifurcation diagrams shown in figure \ref{simutemp_nfp} (left), figure \ref{simutemp_nfp} (right) shows results of time-domain simulations. The first simulation (left) corresponds to the blue point on the ghost note branch of the bifurcation diagrams. Its instantaneous frequency is $\SI{69.5}{\hertz}$, which is coherent with the oscillation frequency displayed on the top right plot in figure \ref{simutemp_nfp}. The second time-domain simulation (right) corresponds to the orange point on the pedal note branch of the bifurcation diagrams. Its instantaneous frequency is $\SI{60.6}{\hertz}$, which is coherent with the oscillation frequency displayed on the top left bottom in figure \ref{simutemp_nfp}.

In the case of the pedal note, it is worth noting that over one period, $p(t)$ becomes higher than $p_m$ (horizontal dotted line), which is physically questionable as it was never observed experimentally. This phenomenon might be considered as a weakness of the model used, caused by the raw modelling of the contact between the lips in equation \ref{equ}. To overcome this limitation, a contact force between the lips -- currently unavailable in literature -- could be added in this equation. However, it is worth mentioning that such a contact force has already been introduced in the case of reed instruments \cite{munozaranconVivoVitroCharacterization2013a}.

\subsection{Assessment of the frequency interval between pedal note and ghost note}

\begin{figure}[h!]
    \centering
    \includegraphics[width=\linewidth]{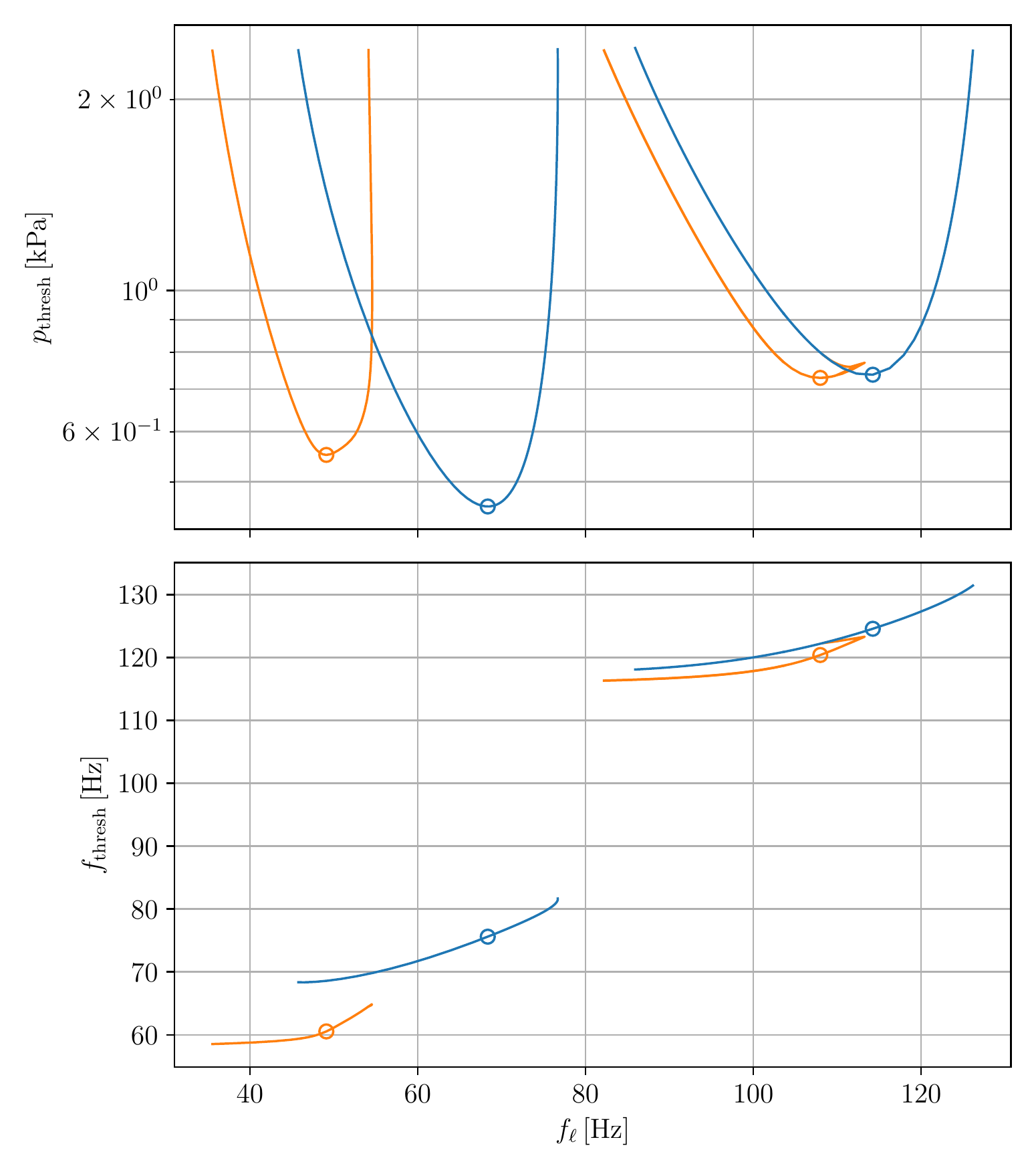}
    \caption{\baselineskip 3mm Top and bottom plots represent respectively the threshold pressures and threshold frequencies given by the linear stability analysis (blue) and the continuation method (orange) vs. the lips' resonance frequency in the case of a euphonium. The circles identify the value of $\fl$ corresponding to a local minimum of blowing pressure. In the case of the second regime, the easiest note playable is given by the minimum of the continuation method (orange curve), since its minimal threshold mouth pressure is lower than the one given by the linear stability analysis (blue curve). The first two regimes on the left of each plot represent respectively the pedal note and the ghost note.}
    \vspace{0mm}
    \label{psc_vs_asl_euphonium_zoom12}
\end{figure}

It is possible to qualitatively check the relevance, in terms of intonation, of these two regimes predicted  by the bifurcation analysis. In particular, one can compare the frequency intervals between the \gu{easiest-to-play} notes predicted by the bifurcation diagrams on the one hand, and by analysing recordings of a euphonium player on the other hand. In the case of a euphonium and considering the U-shaped patterns in figure \ref{psc_vs_asl_euphonium_zoom12}, the \gu{easiest-to-play} notes are those with the lowest threshold mouth pressure. The corresponding oscillation frequencies are $\fson{per}{\text{PN}} = \SI{60.6}{\hertz}$ for the pedal note, $\fson{eq}{\text{GN}} = \SI{75.6}{\hertz}$ for the ghost note, and $\fson{per}{2} = \SI{120.4}{\hertz}$ for the B$\flat$2 (in the bifurcation diagrams, we have actually $\fson{per}{2} < \fson{eq}{2}$). Therefore, it has been found that the frequency interval between the pedal note and the ghost note inferred from the bifurcation analysis ($75.6/60.6 \approx 1.25 = 5/4$) is about \SI{2}{tones} (major third), whereas the recording of a euphonium player\footnote{\label{tubarec}Recording of a euphonium player at \url{http://perso.univ-lemans.fr/~rmatte/bb-euphonium_reg1-6.mp4} and of an E$\flat$-tuba player at \url{http://perso.univ-lemans.fr/~rmatte/eb-tuba_reg1-6.mp4} playing successively an ascending series and a descending series of the six first natural notes, going through the ghost note when descending.} leads to a high major third between the pedal note and the ghost note, which is reasonably close. Furthermore, the frequency interval between the pedal note and the second regime obtained in the bifurcation diagrams is found to be close to \SI{12}{semitones} (octave), as one would expect between a B$\flat$1 and a B$\flat$2.

Eventually, it is worth highlighting that the ghost note does exist for other tuba types than the euphonium, such as the E$\flat$-tuba for instance\textsuperscript{\ref{tubarec}}.

\section{Conclusion}

Most results in this study highlight the usefulness of linear stability analysis to understand various near-threshold behaviours of a complete nonlinear model of brass instrument applied to trombone and euphonium, as well as the ability of bifurcation diagrams to quickly give valuable information about their periodic solutions.

The ease of playing of an instrument is chosen to be assessed based on the minimal threshold pressures. A good estimate can be obtained using linear stability analysis. At the same time, the associated threshold oscillation frequency is inferred, which is directly linked to the pitch of the note. For the considered elementary model of brass instrument, this method provides information on all the periodic regimes accessible to a trombone player for a given position of the slide, and on almost all periodic regimes accessible for a given fingering of a euphonium. 
 
 However, linear stability analysis turns out to be unable to describe the pedal note of the euphonium. In this respect, the analysis of bifurcation diagrams are conveniently introduced. It allows to illustrate the existence of the pedal regime in the form of a branch of periodic solutions far from equilibrium (hence the failure of linear stability analysis to describe it) separated from the one describing the ghost note, which itself emerges from the destabilisation of the equilibrium solution. Moreover, compared to the linear stability analysis, bifurcation diagrams give a more precise prediction of the threshold mouth pressures in the event of an inverse Hopf bifurcation. However, it is worth highlighting that the minimal threshold mouth pressure of each periodic regimes for a given slide position or fingering in the case of a trombone or a euphonium -- except the pedal note of the latter -- given by the bifurcation diagrams is qualitatively the same as given by the linear stability analysis.

If the ghost note seems to exist on every model of tubas, the interval between the ghost note and the first regime appears to vary from a minor third to a perfect fourth depending on the tuba's pitch. Future work include a more quantitative study on a wide range of tubas going from the B$\flat$-contrabass tuba to the B$\flat$-baritone horn (an octave higher), in order to determine whether or not the frequency interval between the pedal note and the ghost note remains constant regardless of the considered bass brass instrument with predominantly-expanding bore profile.

Moreover, it is worth noting that the pedal note also exhibits different behaviours for the high brass instruments, depending on the bore profile of the instrument. For example, the pedal note appears to be more \gu{slotted} (meaning that the player has the sensation of a well-defined pitch) on a flugelhorn (predominantly-expanding bore profile) than on a trumpet (predominantly-cylindrical bore profile).

One important limitation of the present study lies in the difficulty to estimate the lips' parameters used in the model. Indeed, it would be reasonable to think that lips' parameters would vary from playing a low note to playing a high note, not to mention from playing a contrabass tuba to playing an alto trombone for instance. Yet, these lips' parameters being difficult to measure both on an artificial mouth or directly on the player, they were kept the same regardless the instrument or the note played. Even though the obtained results look reasonable, i.e consistent with the musicians’ experience, \emph{in vivo} measurements of lips parameters during musical performance would be very valuable. Furthermore, even if the system dynamics does not seem to depend sensitively on the model's parameters such as the lip-opening height at rest $H$, it may be worth checking the impact of a finer modelling of various elements on the ease of playing and pitch: for instance, considering a mouth section proportional to $H^2$ (instead of only $H$ in the present study), taking a second resonance frequency of the lips into account, or adding a contact force between the lips when they are about to touch each other.

\section{Acknowledgements}

We wish to thank Emmanuel Brasseur and Julia Mourier for useful discussions and for their help in this work.

\begin{appendix}

\section{Inverse Fourier transform of the input impedance}
\label{IFTII}

This appendix is based on the demonstration found in section 6.3 of \cite{silvaEmergenceAutooscillationsDans2009}.

Quantities in the Fourier domain are written using capital letters, whereas quantities in the time domain are written using small letters.

Starting from equation (\ref{ZC}), we have $P(\w) = \Zr(\w) U(\w)$ by definition of the input impedance. Since $\Zr$ represents the impulse response of the resonator, the inverse Fourier transform of $\Zr$ has to be real. Thus, $\Zr$ is hermitian-symmetric, that is to say $\Zr(-\w) = \Zr^*(\w)$. The pressure in the mouthpiece in the time domain $p$ can then be inferred by taking the inverse Fourier transform of $P$:

\[
    \begin{split}
        p(t) &= \iTF{P(\w)}(t)\\
             &= \frac{1}{2\pi} \int_{\R} \Zr(\w) U(\w) \e^{\J\w t} \diff\w\\
             &= \frac{1}{2\pi} \sum_{n=1}^N \bigg(\int_{\R} \frac{C_n}{\J\w-s_n} U(\w) \e^{\J\w t} \diff\w\\
             &\hspace{5mm}+ \int_{\R} \frac{C_n^*}{\J\w-s_n^*} U(\w) \e^{\J\w t} \diff\w\bigg)\\
             &= \frac{1}{2\pi} \sum_{n=1}^N \bigg(\int_{\R} \frac{C_n}{\J\w-s_n} U(\w) \e^{\J\w t} \diff\w\\
             &\hspace{5mm}+ \int_{\R} \frac{C_n^*}{-\J\w'-s_n^*} U(-\w') \e^{-\J\w't} \diff\w'\bigg),
    \end{split}
\]

\nin with $\w' = -\w$. $U$ being also hermitian-symmetric, we have:

\[
    \begin{split}
        p(t) &= \frac{1}{2\pi} \sum_{n=1}^N \Bigg(\int_{\R} \frac{C_n}{\J\w-s_n} U(\w) \e^{\J\w t} \diff\w\\
             &\hspace{5mm}+ \int_{\R} \pr{\frac{C_n}{\J\w'-s_n} U(\w') \e^{\J\w't}}^* \diff\w'\Bigg)\\
             &= \frac{1}{2\pi} \sum_{n=1}^N 2\Re\pr{\int_{\R} \frac{C_n}{\J\w-s_n} U(\w) \e^{\J\w t} \diff\w}\\
             &= \frac{1}{2\pi} \sum_{n=1}^N 2\Re\pr{p_n(t)},
    \end{split}
\]

\nin where $p_n$ is the complex component of the pressure in the mouthpiece $p$, whose dynamical behaviour is linked to $C_n$ and $s_n$:

\[
    P_n(\w) = \frac{C_n}{\J\w-s_n} U(\w) \Rightarrow \dot{p}_n(t) - s_n p_n(t) = C_n u(t).
\]

\section{Dimensionless variables and equations}
\label{DVEUPP}

From a numerical point of view, it is reasonable to make equations dimensionless in order to prevent the variables from varying on a too-large scale, especially when using the continuation method in AUTO. In all that follows, most dimensionless quantity are written with a tilde \gu{$\sim$} unless otherwise stated.

The dimensionless variables of the system $\pr{\ad{h},\ad{p},\ad{u}}$, the dimensionless time $\ad{t}$, the dimensionless mouth pressure $\gamma$, the dimensionless lips' resonance frequency $\thl$ and the dimensionless modal parameters $\pr{\kappa_n,\sigma_n}$ are defined as follows :

\[
    \ad{h} = \frac{h}{h_0} \;;\; \ad{p} = \frac{p}{p_0} \;;\; \ad{u} = \frac{Z_c}{p_0} u \;;
\]
\[
    \ad{t} = \w_0 t \;;\; \gamma = \frac{p_m}{p_0} \;;\; \thl = \frac{\wl}{\w_0} \;;
\]
\[
    \kappa_n = \kap{n} \;;\; \sigma_n = \sig{n},
\]

\nin
where $h_0$ is given in table \ref{lips_params}, $\w_0 = \Im\pr{s_4}/4$ is the quarter of the fourth resonance angular frequency of the resonator (chosen as so because the fourth resonance frequency appears to be quite similar between instruments of the same tube length, such as a trombone and a euphonium for instance), so that $p_0$ can be defined in a similar way as the closure pressure for woodwind instruments \cite{colinotInfluenceGhostReed2019}: $p_0 = \mu \w_0^2 h_0$. Given the new set of variables, the state vector is now:

\[
    \begin{split}
        \ad{\X} &= \pr{\br{\ad{X}_m}_{m \in \dhk{1,2(N+1)}}}\\
                &= \pr{ \ad{h}; \ad{\dot{h}}; \br{\Re(\ad{p}_n)}_{n \in \dhk{1,N}}; \br{\Im(\ad{p}_n)}_{n \in \dhk{1,N}}},
    \end{split}
\]

\nin
so the equation (\ref{sys}) is now written as:

\[
    \ad{F}:\ad{\X}\mapsto
    \begin{pmatrix}
        \ad{X}_2\\
        - \agaml \ad{X}_2 - \thl^2\pr{\ad{X}_1-\ad{H}} + \gamma-2\sum_{n=3}^{N+2}\ad{X}_n\\
        \Re\hk{\sigma_1\pr{\ad{X}_3 + \J \ad{X}_{N+3}} + \kappa_1 \ad{u}(\ad{\X})}\\
        \Re\hk{\sigma_2\pr{\ad{X}_4 + \J \ad{X}_{N+4}} + \kappa_2 \ad{u}(\ad{\X})}\\
        \vdots\\
        \Re\hk{\sigma_N\pr{\ad{X}_{N+2} + \J \ad{X}_{2(N+1)}} + \kappa_N \ad{u}(\ad{\X})}\\
        \Im\hk{\sigma_1\pr{\ad{X}_3 + \J \ad{X}_{N+3}} + \kappa_1 \ad{u}(\ad{\X})}\\
        \Im\hk{\sigma_2\pr{\ad{X}_4 + \J \ad{X}_{N+4}} + \kappa_2 \ad{u}(\ad{\X})}\\
        \vdots\\
        \Im\hk{\sigma_N\pr{\ad{X}_{N+2} + \J \ad{X}_{2(N+1)}} + \kappa_N \ad{u}(\ad{\X})}\\
    \end{pmatrix},
\]

\nin
with $\ad{u}$ now written as:

\[
    \ad{u}:\ad{\X} \mapsto \zeta \times \frac{\ad{X}_1 + \sqrt{\ad{X}_1^2+\eta}}{2} \times \frac{\pr{\gamma-2\sum_{n=3}^{N+2}\ad{X}_n}}{\sqrt[4]{\pr{\gamma-2\sum_{n=3}^{N+2}\ad{X}_n}^2+\eta}},
\]

\nin
and eventually $\zeta = \frac{w Z_c}{\w_0} \sqrt{\frac{2h_0}{\varrho \mu}}$ the dimensionless lip-opening height at rest, defined similarly to \cite{gilbertMinimalBlowingPressure2020a}. 
\end{appendix}

\small
\bibliographystyle{apalike}
\bibliography{brass}

\end{document}